\documentclass[12pt,oneside,letterpaper]{article}
\usepackage{titlesec}
\titleformat*{\section}{\large\bfseries}
\titleformat*{\subsection}{\normalsize\bfseries}
\usepackage{amssymb}
\usepackage{amsmath}
\usepackage[dvips]{graphicx}
\usepackage{setspace}
\usepackage{stackrel}
\usepackage{amsfonts}
\usepackage{fancyhdr}
\usepackage[dvipsnames]{xcolor}
\usepackage{caption}
\usepackage{graphicx}
\usepackage{rotating}
\usepackage{comment}
\usepackage{color}
\usepackage{cite}
\usepackage{braket}

\usepackage{subfig}

\definecolor{darkgreen}{rgb}{0,0.5,0}
\definecolor{darkblue}{rgb}{0,0,0.6}
\definecolor{purple}{rgb}{0.4,.2,0.7}
\newcommand{\p}{\partial}

\newcommand{\f}{\frac}

\newcommand{\be}{\begin{equation}}
\newcommand{\ee}{\end{equation}}

\newcommand{\mc}{\mathcal}

\definecolor{penroseblue}{RGB}{94,153,220}
\definecolor{penrosered}{RGB}{158,28,74}

\usepackage{tikz}
\usetikzlibrary{matrix}
\usetikzlibrary{decorations.markings,calc,shapes,decorations.pathmorphing,patterns,decorations.pathreplacing}
\usetikzlibrary{positioning}

\usepackage[colorlinks=true,citecolor=darkgreen,linkcolor=black,urlcolor=purple]{hyperref}

\usepackage{pdfsync}

\makeatletter
\newcommand*{\defeq}{\mathrel{\rlap{%
      \raisebox{0.3ex}{$\m@th\cdot$}}%
      \raisebox{-0.3ex}{$\m@th\cdot$}}%
      =} 
\makeatother

\DeclareMathOperator{\Tr}{Tr}
\def\be{\begin{eqnarray}}
\def\ee{\end{eqnarray}}

\newcommand{\bea}{\begin{eqnarray}}
\newcommand{\eea}{\end{eqnarray}}

\newcommand{\igapless}{I_{\text{gapless}}}

\def\ben{\begin{equation}}
\def\een{\end{equation}}

\let\a=\alpha \let\b=\beta  \let\d=\delta 
  \let\q=\theta 
\let\l=\lambda \let\m=\mu \let\n=\nu  \let\p=\phi \let\r=v
 \let\t=\tau

    \let\L=\Lambda

\let\f=\frac

\def\be{\begin{equation}}
\def\ee{\end{equation}}
\def\ba{\begin{array}}
\def\ea{\end{array}}

\def\bt{\tilde{\b}}

\def\ba#1\ea{\begin{align}#1\end{align}}
\def\bs#1\es{\begin{split}#1\end{split}}

\usepackage{amsmath}

\renewcommand{\p}{\partial}

\usepackage[top=1in, bottom = 1in, left = 1in, right = 1in]{geometry}

\allowdisplaybreaks 

\begin{document}
\onehalfspacing

\begin{center}

~
\vskip5mm

{\LARGE {
Flat space gravity\\ \vspace{2mm} at finite cutoff
}}

\vskip8mm

Batoul Banihashemi, Edgar Shaghoulian, and Sanjit Shashi*

\vskip8mm
{ \it UC Santa Cruz\\
Physics Department\\
1156 High Street\\
Santa Cruz, CA 95064}


\end{center}

\vspace{4mm}

\begin{abstract}
\noindent
We study the thermodynamics of Einstein gravity with vanishing cosmological constant subjected to conformal boundary conditions. Our focus is on comparing the series of subextensive terms to predictions from thermal effective field theory, with which we find agreement for the boundary theory on a spatial sphere, hyperbolic space, and flat space. We calculate the leading Wilson coefficients and observe that the first subextensive correction to the free energy is negative. This violates a conjectured bound on this coefficient in quantum field theory, which we interpret as a signal that gravity does not fully decouple in the putative boundary dual.
\end{abstract}
\vfill



\thispagestyle{empty}
\pagebreak
\pagestyle{plain}

\setcounter{tocdepth}{2}
{}
\vfill
\clearpage
\setcounter{page}{1}

\tableofcontents
\vfill\pagebreak

\section{Introduction}
The holographic principle \cite{tHooft:1993dmi,Susskind:1994vu} purports to equate quantum gravity in spacetimes with a given boundary structure to a lower-dimensional theory located at the boundary. The best-understood example of this is the family of dualities relating supersymmetric Yang-Mills theory to geometries sourced by stacks of D-branes. This family includes the famous AdS/CFT correspondence. Part of the power of these examples is that the boundary dual is a local QFT for $d \leq 5$. A flurry of recent work has motivated studying the structure of finite timelike boundaries in such contexts \cite{McGough:2016lol, Taylor:2018xcy, Hartman:2018tkw, Gorbenko:2018oov, Gross:2019ach, Lewkowycz:2019xse, Gross:2019uxi, Coleman:2020jte, Coleman:2021nor, Svesko:2022txo, Banihashemi:2022jys, Anninos:2023epi, Anninos:2024wpy, Batra:2024kjl, Batra:2024qju}. When trying to extend gauge/gravity duality to more general spacetimes, however, one faces the feature that the simplest observables---when translated from $(d+1)$-dimensional bulk variables to $d$-dimensional boundary variables under the assumption of the usual holographic dictionary---do not reflect a local QFT structure. 

In remarkable recent developments \cite{Anninos:2023epi, Anninos:2024wpy}, it was shown that this conclusion could be modified by considering novel boundary conditions in the gravitational path integral. These boundary conditions are the so-called ``conformal boundary conditions," where one fixes the conformal class of the metric and the trace of the extrinsic curvature $K$. This is to be contrasted with Dirichlet boundary conditions, where one fixes the metric entirely. Recent work \cite{Anderson:2006lqb,An:2021fcq} suggests that conformal boundary conditions may be well-posed in situations where the Dirichlet conditions are not. The authors of \cite{Anninos:2023epi, Anninos:2024wpy} studied horizon thermodynamics with conformal boundary conditions and found that entropy behaves as $S \sim \f{\text{Vol}}{GK^{d-1}} \,T^{d-1}$ at high temperature, precisely as would be expected for a local boundary dual. 

In this paper we will focus on the case of gravity with zero cosmological constant. Our goal is to compare the bulk gravitational calculations to more precise tests of locality in a putative boundary theory. In particular, we will compare to the boundary theory's thermal effective action, which makes a prediction for the series of subextensive corrections to the entropy on arbitrary spatial backgrounds. 
In all considered cases we will find agreement between the bulk theory and the thermal effective action, providing further evidence that the boundary dual theory is a local theory, at least in the high-temperature limit where the cutoff surface goes to infinity. However, we will find that the leading subextensive correction to the free energy $I = -\log Z$ is \emph{negative}, violating a conjectured QFT bound from \cite{allameh}. We will interpret this as evidence that gravity does not fully decouple in the boundary dual theory. 

In the following subsections, we will review conformal boundary conditions in gravity and the machinery of the thermal effective action, setting the stage for the rest of the paper.

\subsection{Conformal boundary conditions}

A typical choice of boundary conditions in any theory with boundary is the Dirichlet condition fixing the fields. In Einstein gravity with zero cosmological constant, this would mean solving the following boundary value problem on the manifold $\mathcal{M}$:
\begin{equation}
G_{\mu\nu} = 0,\quad \delta h_{\mu\nu} = 0.\label{bcproblemDirichlet}
\end{equation}
Here, $G_{\mu\nu}$ is the bulk Einstein tensor, and $h_{\mu\nu}$ is the induced metric on $\partial\mathcal{M}$. The corresponding Euclidean action is 
\begin{equation}
\text{Dirichlet boundary conditions:}\quad I = -\frac{1}{16\pi G}\int_{\mathcal{M}} d^{d+1}x \sqrt{g}\,R - \frac{1}{8\pi G}\int_{\partial\mathcal{M}}d^d x \sqrt{h}\,K.
\end{equation}
However, the initial boundary value problem with Dirichlet conditions is generally ill-posed,\footnote{More specifically, the constraint equation projected on the boundary is not satisfied by generic Dirichlet data. In addition to the existence problem there are certain cases, such as flat boundaries in both Euclidean \cite{Anderson:2006lqb, Witten:2018lgb} and Lorentzian \cite{An:2021fcq, Anninos:2022ujl} signature, where uniqueness is violated.} and it is conjectured that ``conformal" boundary conditions are well-posed \cite{Anderson:2006lqb,Witten:2018lgb,An:2021fcq}. So rather than freezing the boundary metric completely, we only fix it up to an overall conformal mode and in conjunction fix the trace of the extrinsic curvature. So instead of \eqref{bcproblemDirichlet}, we solve
\begin{equation}
G_{\mu\nu} = 0,\quad \delta(h^{-1/d}h_{\mu\nu}) = 0,\quad \delta K = 0.
\end{equation}
We have introduced the conformal metric $\overline{h}_{\m\n} = h^{-1/d} h_{\m\n}$, which is invariant under Weyl rescalings of $h_{\m\n}$. We stress that this conformal metric is a tensor \emph{density} due to the $h^{-1/d}$ factor, which transforms with weight $-2/d$. For a study of these boundary conditions in the fluid-gravity literature see \cite{Bredberg:2011xw, Anninos:2011zn}.

The Euclidean action consistent with conformal boundary conditions is \cite{York:1972sj,York:1986lje,Odak:2021axr}
\begin{equation} \label{I_CBC}
\text{Conformal boundary conditions:}\quad I = -\frac{1}{16\pi G}\int_{\mathcal{M}} d^{d+1}x\sqrt{g}\,R - \frac{1}{8d\pi G}\int_{\partial\mathcal{M}}\hspace{-2mm}d^{d}x \sqrt{h}\,K.
\end{equation}
The Brown--York stress tensor computed from this action, $T^{\text{CBC}}_{\m\n} = -(K_{\m\n} - \frac{1}{d}h_{\m\n}K)$, is manifestly traceless. We will occasionally use the CBC shorthand to highlight the assumption of conformal boundary conditions. The energy computed from this stress tensor agrees with the boundary value of the Hamiltonian derived from this action, as shown in Appendix \ref{appA}.

Taking the hypothesis that the bulk theory with these boundary conditions can be captured by a putative boundary dual theory, we examine some features of this dual theory through the \textit{thermal effective action}. We will always take $K$ to be a constant, although in general it can depend on the boundary spacetime coordinates. We will find that we can generically accommodate any $K > 0$, but that in some cases we can have $K \leq 0$ as well. 

\subsection{Thermal effective theory}\label{subsec:teftintro}
Let us consider a field theory on a spatial $(d-1)$-dimensional manifold $\Sigma$ ($d \geq 3$) and its canonical partition function at finite temperature $1/\b$. At high temperature, we can dimensionally reduce over the thermal circle $S^1_\b$ and write an effective action for the gapped sector as \cite{Jensen:2012jh, Banerjee:2012iz}
\be
-\log Z(\b) = \int_{\Sigma} d^{d-1} x\,\sqrt{h_{\Sigma}} \left(-\f{c_0}{\b^{d-1}} + \f{c_1}{\b^{d-3}} R +\cdots\right) + \igapless + I_{\text{np}},\label{teftact}
\ee
where $R$ is the Ricci scalar of $\Sigma$ and $I_{\text{np}}$ describes nonperturbative corrections. For generic interacting theories, $\igapless = 0$ and the equilibrium thermodynamics is captured by the effective action of background terms \cite{Banerjee:2012iz,horowitz,Benjamin:2023qsc}. An example is holographic theories dual to gravity in AdS spacetime \cite{horowitz,Benjamin:2023qsc,allameh}.

The coefficient $c_0$ determines the extensive high-temperature entropy, which implies that $c_0 > 0$. Meanwhile, the series predicted by the thermal effective action is a refined test of locality and Lorentz invariance that goes beyond the leading extensive term. For example, there is a conjectured bound on the first subleading correction $c_1 \geq 0$ \cite{allameh} motivated by results in both free theories \cite{Kutasov:2000td, Melia:2020pzd, Benjamin:2023qsc} and conformal field theories dual to AdS gravity with a Dirichlet boundary \cite{allameh}.

While \cite{Anninos:2023epi} showed consistency of the leading extensive piece of \eqref{teftact} with flat-space gravity with conformal boundary conditions, in this paper we would like to investigate the consistency of the subleading terms. This amounts to checking the scaling of the subleading terms with $\b$ and their dependence on the spatial geometry. Before we do this, let us discuss how the thermal effective theory interfaces with conformal boundary conditions. In particular since the induced metric is not fixed on the boundary we need to understand what becomes of the curvature invariants and the determinant of the metric $h_\Sigma$.

\subsection{Bulk-boundary dictionary}\label{bbdrydic}

To apply the thermal effective action framework to gravity with conformal boundary conditions, we will assume a single, primitive entry of the bulk-boundary dictionary: the equality of partition functions. 
The partition function of the bulk gravitational theory is presumably given by a path integral over metrics (with conformal boundary conditions) and any additional fields. We will evaluate this path integral by a saddlepoint approximation throughout this paper. Thus to leading order the partition function is estimated via the on-shell action of the Euclidean saddlepoint, computed using \eqref{I_CBC}: $Z \approx \exp(-I)$. While the putative boundary theory is unknown to us, we assume it also has a well-defined path integral that defines its partition function. 

We will consider the boundary theory defined on a class of geometries conformal to $S^1 \times \Sigma_{d-1}$,
\begin{equation}
h_{\mu\nu}dx^\m dx^\n = \l^2\, \tilde{h}_{\m\n} dx^\m dx^\n = \l^2 \,(d\tilde{\tau}^2 + d\Sigma_{d-1}^2),\qquad \tilde{\t} \sim \tilde{\t} + \tilde{\b},\label{confrep}
\end{equation}
where we have picked a special representative $\tilde{h}_{\m\n}$. $d\Sigma_{d-1}^2$ is the line element on the spatial manifold $\Sigma_{d-1}$. The representative $\tilde{h}_{\m\n}$ is the same as the conformal metric $\bar{h}_{\m\n}$ in the special case that $d\Sigma_{d-1}^2 = dx_i^2$. We will often consider $\Sigma_{d-1} = S^{d-1}$ with $d\Sigma_{d-1}^2$ set to the round metric $d\Omega_{d-1}^2$ on the unit sphere. The quantity $\tilde{\beta}$ is then the inverse ``conformal" temperature, which in any representative of the conformal class with constant $\l$ is the ratio of the proper inverse temperature to the proper radius of the sphere. This characterization of $\tilde{\b}$ has to be modified in a general representative where $\l$ can depend on the coordinates, but this subtlety will not play a role for us.\footnote{Importantly, if for example the Weyl symmetry of the boundary is to be thought of as a gauge symmetry, then calculations phrased in terms of $\tilde{\b}$ need to be interpreted with care. In particular, if we want to conclude that extensive high-temperature entropy implies locality, we would have to assume that it is sufficient to exhibit extensivity in the special representative $\tilde{h}$ that we have chosen.}

We will make two assumptions. First, we assume that the boundary effective action is written in terms of curvature invariants of the chosen representative $\tilde{h}_{\m\n}$.\footnote{If we restrict to a conformal class of metrics with constant conformal mode $\l$, then we can write an effective action with the geometric quantities referring to any member of the conformal class $\{h_{\m\n}\}$, e.g. $\b = \l \tilde{\b}$ and the Ricci scalar of the full spatial geometry $\l^2 d\Sigma_{d-1}^2$. The result will be independent of $\l$.} Second, we treat $K$ as a parameter of the theory and \textit{not} as an additional background term that can explicitly appear in the effective action. Altogether, we assume the following thermal effective action:
\be\label{flateft}
-\log Z(\tilde{\b}) = \int d^{d-1}x \sqrt{\tilde{h}_\Sigma} \left(-\f{c_0}{\tilde{\b}^{d-1}} + \f{c_1}{\tilde{\b}^{d-3}}\tilde{R}+\cdots\right).
\ee
$\tilde{R}$ is the Ricci scalar of $\Sigma_{d-1}$ and $\tilde{h}_\Sigma$ is the determinant of its induced metric. This is the proposal we would like to test, and indeed we will find precise agreement with the gravitational calculation. Interestingly, however, we find that the first subleading correction to the free energy has a coefficient $c_1$ that is \emph{negative}. This violates the conjectured QFT bound $c_1 \geq 0$ from \cite{allameh}. We will interpret this negativity as a signal that gravity does not fully decouple in the boundary dual.

We would also like to have a proposed Hilbert-space picture for the partition function. In an ordinary QFT, with temperature as our only potential, we have $Z(\b) = \Tr \exp[-\b H]$. For our putative boundary theory, we will use the Hamiltonian $H_{\rm p}$ that generates the boundary proper time $\l \tilde{\t}$, where we have put the boundary metric in the form \eqref{confrep}. This Hamiltonian is conjugate to the periodicity $\b_{\rm p} = \l \tilde{\b}$ of the proper time coordinate and gives $Z = \Tr \, \exp[-\b_{\rm p} H_{\rm p}]$.

For conformal boundary conditions we have $H_{\rm p} = H^{\rm CBC}$, whose value is given by (see Appendix \ref{appA} for the derivation)
\be\label{E-CBC}
E^{\rm CBC}= \int \! d^{d-1}x \, \sqrt{\sigma_{d-1}} \, u^\mu u^\nu T_{\mu\nu}^{\rm CBC}= - \frac{\text{Vol}[S^{d-1}]}{8\pi G} \frac{d-1}{d}r_c^{d-2}\sqrt{f(r_c)}\left (1-\frac{r_c}{2}\frac{f'(r_c)}{f(r_c)} \right).
\ee
We can therefore write 
\be\label{Z-conformal}
Z = \Tr \exp[-\b_{\rm p} H_{\rm p}] = \Tr \exp[-\tilde{\b} \tilde{H}] ,
\ee
where we have defined $\tilde{H} \equiv \l H^{\rm CBC}$ as the generator of time translations in $\tilde{\t}$.

\subsection{Outline}

In Section \ref{sec:fecbc}, we examine the solution space of flat-space gravity with an $S^1 \times S^{d-1}$ boundary. We also extract the Wilson coefficients of the thermal effective action.

In Section \ref{sec:nonsphere}, we extend the analysis of flat-space gravity to boundaries whose spatial manifolds are not spherical, but rather locally flat or hyperbolic. We confirm that the Wilson coefficients are the same as those obtained in Section \ref{sec:fecbc}. However, the solution spaces in these cases are quite different. A flat boundary yields bulk solutions that are simply compactifications of Rindler space, whereas a hyperbolic boundary yields a ``cosmic" solution with a valid high-temperature limit only for $K > 0$.

In Section \ref{sec:thermo1st}, we evaluate the first law of thermodynamics with conformal boundary conditions, so as to better understand the role of $K$ in a putative boundary theory. We find that $K$ appears thermodynamically conjugate to the spacetime volume of the boundary. We differentiate this from pressure, which appears conjugate to a spatial volume, and show how to recover ordinary pressure-volume terms in black hole thermodynamics.

In Section \ref{sec:spherePart}, we compute the sphere partition function by solving for bulk solutions with boundary $S^d$, again motivated by understanding the role of $K$. We find that the result is consistent with the tracelessness of the boundary stress tensor.

In Section \ref{sec:bdryCond}, we explore the universality of the extensive scaling in the thermal effective action. For conformal boundary conditions, we show that extensivity is guaranteed in a wide class of examples due to an argument involving the universality of the Rindler approximation to finite-temperature horizons. In this sense, conformal boundary conditions are special. Others, such as the one-parameter family of well-posed boundary conditions discussed by \cite{Liu:2024ymn}, exhibit non-extensive scaling.

We end with some concluding remarks and outlook in Section \ref{conclusions}.

\section{Thermodynamics in the conformal canonical ensemble}\label{sec:fecbc}

We now calculate the partition function in the ``conformal" canonical ensemble by a saddle point approximation of the gravity path integral in the bulk. To reiterate, our goal is to show that this reproduces the thermal expansion \eqref{flateft}. We also seek to calculate the Wilson coefficients $c_0$ and $c_1$. To do so, we will consider Einstein gravity with conformal boundary conditions and spatial manifold $\Sigma_{d-1} = S^{d-1}$.

Furthermore, we assume the bulk geometry is spherically symmetric,\footnote{While $\tilde{h}$ is spherically symmetric, this symmetry is only a conformal isometry of an arbitrary representative of the conformal class. Thus, it may be more natural to assume that the bulk geometry only preserves conformal Killing vectors which form the algebra of $SO(d)$. We are unaware of any black hole uniqueness theorems under the assumption of conformal Killing vectors instead of ordinary Killing vectors.} in which case Birkhoff's theorem tells us the Schwarzschild metric provides the full space of solutions for Einstein gravity with zero cosmological constant. This is a particularly powerful statement, since it is valid within any finite region; if we eliminate the assumption of spherical symmetry, then we can construct static solutions of the vacuum Einstein equations by introducing stress-energy ``beyond the cutoff." For a large class of (Lorentzian) axisymmetric black holes constructed by introducing external stress-energy see \cite{Geroch:1982bv}.\footnote{We thank Gary Horowitz for bringing this paper to our attention.}

\subsection{Bulk solution space}\label{subsec:solsConfCanon}

Conformal boundary conditions require that we fix the conformal class of the metric and the trace of the extrinsic curvature $K$. Fixing the conformal class is equivalent to fixing the conformal metric $\bar{h}_{\m\n} = h^{-1/d}h_{\m\n}$. To have a good variational principle requires the boundary term in \eqref{I_CBC}. We work in Euclidean signature and choose the conformal class of metrics whose representatives can be written as 
\be \label{sphr-bdry}
\left.ds^2\right|_{\partial\mathcal{M}} = \l^2\left(d\tilde{\tau}^2 + d\Omega_{d-1}^2\right),\quad \tilde{\tau} \sim \tilde{\tau} + \tilde{\b},
\ee
where $\l$ is an arbitrary function of the coordinates. We will be interested in the dependence of the free energy on the inverse conformal temperature $\tilde{\b}$, discussed around \eqref{confrep}. All solutions we consider will fit into this class with the conformal mode given by $\l = r_c$ for the Schwarzschild radial coordinate $r$ evaluated at the cutoff location. 

Given the assumption of spherical symmetry in the bulk, our exercise now is simply to fit the Schwarzschild solutions into the boundary conditions \eqref{sphr-bdry} and fixed $K$. The zero-mass Schwarzschild is given by
\be\label{thermalgas-0}
ds^2 =d\t^2 + dr^2 + r^2 d\Omega_{d-1}^2,\quad \t \sim \t +\b,
\ee
where $\b$ is an arbitrary parameter and $r \geq 0$. Picking a cutoff $r = r_c$ gives
\be\label{K-thermalgas}
K = \frac{d-1}{r_c}.
\ee
Thus, to satisfy our boundary conditions, we simply set $r_c = (d-1)/K$ and $\b = r_c\,\tilde{\b}$.

The black-hole geometries are given by
\be\label{bhflat}
ds^2 = f(r) d\t^2 + \frac{dr^2}{f(r)} + r^2 d\Omega_{d-1}^2, \quad f(r) \equiv 1-\f{\m}{r^{d-2}},
\ee
with $\mu = r_h^{d-2}$ and $r \geq r_h$. Note that this solution only makes sense when there are more than three bulk dimensions, i.e.~when $d>2$. The period of $\t$ is determined by imposing regularity at $r = r_h$, which yields
\begin{equation}
\t \sim \t + \frac{4\pi}{f'(r_h)} = \t + \f{4\pi r_h}{d-2}.
\end{equation}
At a finite cutoff surface $r = r_c$, the induced metric and extrinsic curvature in terms of an arbitrary emblackening factor are given by
\be
\left.ds^2\right|_{\partial\mathcal{M}} = f(r_c)\,d\t^2 + r_c^2\,d\Omega_{d-1}^2,\qquad K = \frac{d-1}{r_c \sqrt{f(r_c)}}\left[f(r_c) + \frac{r_c f'(r_c)}{2(d-1)}\right].\label{indmetk}
\ee
The inverse conformal temperature $\bt$ is given by the ratio of the proper size of the thermal circle to the proper radius of the spatial sphere. To see this, observe that putting the boundary metric in the form \eqref{sphr-bdry} requires scaling out a factor of $r_c^2$ (so $\l^2 = r_c^2$) and defining $\tilde{\t} = \t\sqrt{f(r_c)} /r_c$. The boundary conditions fix $\tilde{\b}$ and $K$ at the cutoff $r_c$, so we have to solve the equations
\be\label{sphereflateqns}
\tilde{\b} = \f{4\pi r_h}{d-2} \frac{\sqrt{f(r_c)}}{r_c},\qquad K = \frac{1}{2r_c \sqrt{f(r_c)}}\left[2(d-1) - d\left(\frac{r_h}{r_c}\right)^{d-2}\right].
\ee
Finding the solutions explicitly by solving for $r_h$ and $r_c$ as functions of $\tilde{\beta}$ and $K$ is intractable for general $d$ (although see \cite{Anninos:2023epi} for the $d = 3$ analysis and Appendix \ref{app:exact5d} for $d = 4$). Nevertheless, we can count the number of solutions through the following argument. We first note that $\tilde{\beta}$ is purely a function of the dimensionless ratio $z \equiv r_h/r_c \in [0,1]$. In fact, $\tilde{\beta} = 0$ corresponds to $z = 0$ and $1$, and there is a single maximum at $z = \left(2/d\right)^{1/(d-2)}$ at which
\begin{equation}
\tilde{\beta}_{\text{max}} = \frac{4\pi}{d-2}\left(\frac{2}{d}\right)^{1/(d-2)} \sqrt{1 - \frac{2}{d}}.\label{betamax}
\end{equation}
Thus for a given $\tilde{\beta}$, we have zero solutions to the $\tilde{\b}$ equation if $\tilde{\beta} > \tilde{\beta}_{\text{max}}$, and two possible solutions if $\tilde{\beta} < \tilde{\beta}_{\text{max}}$. Suppose that we are in the latter case and denote these two solutions as $z = z^{\pm}$, where $z^+ > z^-$. We can then solve for the corresponding respective values of $r_c$ by solving the equation for $K$:
\begin{equation}
r_c^{\pm} \equiv \frac{1}{2K\sqrt{1 - (z^{\pm})^{d-2}}}\left[2(d-1) - d(z^{\pm})^{d-2}\right].
\end{equation}
The large black hole has horizon radius $r_h = r_c^+ z^+$ and the small black hole has radius $r_h = r_c^- z^-$. So, for a given $\tilde{\beta} < \tilde{\beta}_{\text{max}}$ and $K$, there are a total of three Euclidean solutions---the two black holes and the thermal vacuum geometry \eqref{thermalgas-0}, which we will refer to as the thermal gas. For $\tilde{\beta} > \tilde{\beta}_{\text{max}}$ (low temperature), we only have the thermal gas.

\subsubsection*{The Hawking--Page transition}

As in the case of AdS gravity \cite{Hawking:1982dh,Witten:1998zw}, there exists a first-order phase transition in the semiclassical canonical partition function, which we can probe by computing the free energy from the minimum on-shell Euclidean action:
\begin{equation}
F \approx \tilde{\beta}^{-1}\, \text{min}\,I_{\text{on-shell}}.
\end{equation}
The on-shell Euclidean action \eqref{I_CBC} for both the thermal gas \eqref{thermalgas-0} and the black hole \eqref{bhflat} reduces to a boundary term, since the bulk curvature vanishes:
\begin{equation}\label{Icl}
I_{\text{on-shell}} = -\frac{1}{8d\pi G}\int_{\partial\mathcal{M}} \sqrt{h}\,K = -\frac{\tilde{\beta}K}{8d\pi G}\text{Vol}[S^{d-1}]r_c^{d}.
\end{equation}
We can use the equations for $K$---\eqref{K-thermalgas} for the thermal gas and \eqref{sphereflateqns} for the black hole---to write the free energies of the respective phases as
\begin{equation}
\begin{split}
F_{\text{thermal gas}}
&= -\frac{(d-1)^d \text{Vol}[S^{d-1}]}{8d\pi GK^{d-1}},\\
F_{\text{black hole}}
&= -\frac{(d-1)^d\text{Vol}[S^{d-1}]}{8d\pi GK^{d-1}}\left[1 - \frac{d}{2(d-1)}z^{d-2}\right]^d \left(1 - z^{d-2}\right)^{-d/2}.
\end{split}
\end{equation}
The Hawking--Page point corresponds to the value of $\tilde{\beta}$ at which these two quantities match. By solving the equations, we find that
\begin{equation}
F_{\text{thermal gas}} = F_{\text{black hole}} \implies z^{d-2} = \frac{4(d-1)}{d^2}.
\end{equation}
By plugging this into the equation for $\tilde{\beta}$ \eqref{sphereflateqns}, we find that
\begin{equation}
\tilde{\beta}_{\text{HP}} = \frac{4\pi}{d}\left[\frac{4(d-1)}{d^2}\right]^{1/(d-2)}.
\end{equation}
Plugging in $d = 3$ yields the value $\tilde{\beta}_{\text{HP}} = 32\pi/27$ found by \cite{Anninos:2023epi}, while plugging in $d = 4$ gives us $\tilde{\beta}_{\text{HP}} = \sqrt{3}\pi/2$ as found in Appendix \ref{app:exact5d}. For $z>[4(d-1)/d^2]^{\f{1}{d-2}}$ we have that the large black hole dominates. To see that it is the large one simply requires noticing that this value of $z$ is larger than the value of $z$ corresponding to $\tilde{\b}_{\text{max}}$. (As usual, the small black hole has negative specific heat---defined as $-\tilde{\b}^{2} \p_{\tilde{\b}} \tilde{E}$ at constant $K$---since its temperature increases as it gets smaller and its mass decreases.)

\subsection{Expansion around high temperature}\label{subsec:thermaleftschw}

Let us now calculate the on-shell action in the bulk at high-temperature. We can write the large black-hole solution, which dominates the ensemble as $\tilde{\beta} \to 0$, as a perturbative expansion:
\be \label{expansions}
r_h = \f{2\pi}{\tilde{\b} K}\left(a_0 + a_1 \tilde{\b}^2+a_2 \tilde{\b}^4+\cdots\right),\qquad r_c = \f{2\pi}{\tilde{\b} K}\left(w_0 + w_1 \tilde{\b}^2+w_2 \tilde{\b}^4+\cdots\right).
\ee
This ansatz can be guessed by consistency with the thermal effective action. Plugging this into \eqref{sphereflateqns} and solving for the first few coefficients gives
\begin{align}
\begin{split}
\label{flata1}
a_0 &= 1,\hspace{40mm} w_0 = 1,\\
a_1 &= \f{(d-2)^2}{16\pi^2}, \hspace{27mm}w_1 = \f{(d-1)(d-2)}{16\pi^2},\\
a_2 &= -\f{d(d-2)^2}{256\pi^4},\hspace{22mm} w_2 = \f{(d-1)(d-2)^2}{512 \pi^4},\\
a_3 &= -\frac{d(d+2)(d-2)^3 }{6144 \pi ^6}, \hspace{9.5mm}w_3 = \frac{(d+3)(d-1)(d-2)^3 }{24576 \pi ^6}.
\end{split}
\end{align}
As $R = 0$ for the bulk spacetime, the total on-shell Euclidean action is simply the boundary action, which can be written as in \eqref{Icl}. Plugging in \eqref{expansions} for $r_c$ yields
\begin{equation}
I = {\rm Vol}[S^{d-1}]\left[-\frac{(2\pi)^{d-1}}{4dGK^{d-1}} \frac{w_0^d}{\tilde{\beta}^{d-1}} - \frac{(2\pi)^{d-3}}{GK^{d-1}} \frac{\pi^2 w_0^{d-1} w_1}{\tilde{\beta}^{d-3}} + \cdots\right].\label{genericActExp}
\end{equation}
The scaling of each term in the series with $\tilde{\beta}$ matches the structure of the thermal effective action \eqref{flateft}. In particular the action scales as $1/\tilde{\b}^{d-1}$ at leading order in conformal temperature, which is manifestly extensive.
To read off the Wilson coefficients, we plug in the appropriate values of $w_0$ and $w_1$ in \eqref{flata1} and make use of the fact that $\text{Vol}[S^{d-1}] = \int \sqrt{\tilde{h}_\Sigma}$ in the extensive term and $(d-1)(d-2)\text{Vol}[S^{d-1}] = \int \sqrt{\tilde{h}_\Sigma}\,R$ in the subextensive term. We find that
\be
c_0 = \f{(2\pi)^{d-1}}{4 d G K^{d-1}},\quad c_1 = - \f{(2\pi)^{d-3}}{16GK^{d-1}}.\label{c0c1}
\ee
Interestingly, we have $c_1 < 0$, violating the conjectured bound of \cite{allameh}. As we discuss in Section \ref{conclusions}, we interpret this as a sign that gravity does not fully decouple in the boundary dual theory.

\subsection{Comparison to Dirichlet boundary conditions}

Notice that the equations for $K$ and $\tilde{\b}$ in \eqref{sphereflateqns} can be inverted to write $r_h$ and $r_c$ in terms of $K$ and the proper temperature $\b = r_c \tilde{\beta}$. This means that these solutions can be translated into solutions for Dirichlet boundary conditions at finite cutoff.

When analyzing the high-temperature thermodynamics, one difference between the two sets of boundary conditions is in how the high-temperature limit is taken. For Dirichlet boundary conditions, we tend to take $r_h \rightarrow r_c$ at fixed $r_c$. By contrast, for conformal boundary conditions we take $r_h$ and $r_c$ to infinity with their ratio fixed by $K$. Another important difference is in what parameters we keep fixed in the free energy. Both of these elements lead to the Dirichlet answer being non-extensive, as we now show.

We first pick Dirichlet boundary conditions and take the high-temperature limit $r_h \rightarrow r_c$. The action scales as $I \sim K \b r_c^{d-1}/G$, and in the limit $r_h \rightarrow r_c$ we also have $K \sim \b^{-1}$. Thus, the boundary action scales as $r_c^{d-1}/G$, which is not extensive. Alternatively one can directly use the area law for the entropy, $S \sim r_h^{d-1}/G \sim r_c^{d-1}/G$. 

We can also consider the traditional thermodynamic limit, which is a large-volume limit. In a CFT, this is the same as a high-temperature limit, since the only dimensionless ratio is $\b^{d-1}/{\rm Vol}$. Similarly, for conformal boundary conditions there is no distinction between a large volume or high-temperature limit, since the only dimensionless ratio is $\tilde{\b}$. However, the large-volume limit for Dirichlet boundary conditions is $r_c \rightarrow \infty$ at fixed $\b$. To keep $\b$ fixed, we scale $r_h \rightarrow \infty$. Since $\L = 0$, the only scale besides $\b$ and $r_c$ is $G$, which enters into the thermodynamics as an overall factor. This means that $\b \rightarrow 0$ at fixed $r_c$ and $r_c \rightarrow \infty$ at fixed $\b$ give the same answer for the free energy, which at leading order scales as $r_c^{d-1}/G$. The two limits can be different if there are additional scales; for example, they are different for AdS with Dirichlet boundary conditions at finite cutoff due to the appearance of $\ell_{\text{AdS}}$. 

Now, with the same Dirichlet boundary conditions, let us take the high-temperature limit in the way we would for conformal boundary conditions, keeping fixed
\be
K = \f{1}{2r_c \sqrt{f(r_c)}}\left[2(d-1) - d \left(\f{r_h}{r_c}\right)^{d-2}\right]
\ee
as we take $\b \rightarrow 0$. The fact that $K$ is being fixed is not in the variational sense, it is simply an instruction for how to take the high-temperature limit. Then, the action at leading order is given by 
\be\label{bhdirconf}
I \approx {\rm Vol}[S^{d-1}]\left[-\frac{(2\pi)^{d-1}}{4G}\left(\f{1}{K(r_c,\b)}\right)^{d-1}\frac{r_c^{d-1}}{\b^{d-1}}\right].
\ee
This expression is only interpreted as extensive when the additional factors of $r_c$ and $\b$ are buried in $K$, which is fixed as a boundary condition when fixing conformal boundary conditions, and the rest of the expression is written in terms of $\tilde{\b}$ as $I \sim 1/\tilde{\b}^{d-1}$. But with Dirichlet boundary conditions the fixed parameters are $\b$ and $r_c$, and the answer written in terms of them is not extensive, since $K(r_c,\b)$ will ruin the purely extensive scaling $r_c^{d-1}/\b^{d-1}$.

\section{Boundary theory on non-spherical spaces}\label{sec:nonsphere}

A remarkable aspect of \eqref{flateft} is that, once we have the full set of Wilson coefficients, we can write down the thermal effective action on any spatial manifold simply by calculating the curvatures and doing the integrals. This is difficult in general, although it is quite simple if we restrict to spaces of constant curvature. In this section we will consider hyperbolic space $\mathbb{H}^{d-1}$ and (local) flat space $\mathbb{R}^{d-1}$, the two other maximally symmetric spaces beyond the sphere $S^{d-1}$ which we already considered.

For $\mathbb{R}^{d-1}$, we only have the leading extensive term in the effective action. For $\mathbb{H}^{d-1}$, the result for the free energy is precisely the same as the result for the free energy on $S^{d-1}$, except that every other term (i.e. the terms with an odd power of curvature) has a flipped sign, see \eqref{flateft}.

In the context of AdS/CFT, one can check that this is indeed what happens upon comparing the free energy of hyperbolic black holes to spherical black holes \cite{horowitz}. However, in considering gravity with zero cosmological constant, an immediate obstacle we face in checking the predictions of the thermal effective action is the lack of black hole solutions with nonspherical topology. AdS, on the other hand, has a zoo of such solutions, including planar black branes and hyperbolic black branes (and their compactifications). 

Fortunately, we have a bit of leeway in our problem, since we are working with a finite boundary. The usual theorems regarding nontrivial horizon topology in flat space assume asymptotic flatness, which is rather constraining. We do not assume asymptotic flatness, or asymptotic anythingness. We are therefore able to construct geometries with flat and hyperbolic horizons. While these geometries are not ordinary black holes as in the $S^{d-1}$ case, they are nevertheless valid solutions to the boundary value problem; moreover, they are consistent with the predictions of the thermal effective action. 

\subsection{Solutions with flat horizons}
We will begin by considering our boundary theory on flat space. Our boundary value problem fixes the conformal class of metrics
\begin{equation}
\left.ds^2\right|_{\partial\mathcal{M}} = \l^2\left(d\tilde{\tau}^2 + d\tilde{x}_i^2\right),\quad \tilde{\tau} \sim \tilde{\tau} + \tilde{\beta},\quad \tilde{x}_i \sim \tilde{x}_i + \tilde{L}_i.\label{bdrymetRindler}
\end{equation}
and the extrinsic curvature $K$, which as usual we take to be a constant. So the \textit{conformal} metric on the boundary $h^{-1/d}h_{\mu\nu}$ (as defined in \cite{York:1972sj,York:1986lje}) is locally flat,
and the boundary itself has the topology of a $d$-torus. An unfixed conformal factor implies that it is only the ratios $\tilde{L}_i/\tilde{\beta}$ that are meaningful, which together with $K$ are the $d$ pieces of boundary data.

One geometry which contributes to the gravity path integral with these boundary conditions is a compactification of Rindler space. In Euclidean signature, we can write the metric for this solution as
\begin{equation}
ds^2 = r^2 d\tau^2 + dr^2 + dx_i^2,\quad \tau \sim \tau + 2\pi,\quad x_i \sim x_i + L_i,
\end{equation}
where the $\tau$-cycle caps off at $r = 0$. This compactification of Rindler space violates asymptotic flatness, but we do not preserve this in our problem. At finite cutoff $r = r_c$, we find
\begin{equation}
\left.ds^2\right|_{\partial\mathcal{M}} = r_c^2 d\tau^2 + dx_i^2,\quad K = \frac{1}{r_c}.
\end{equation}
To match the boundary metric in \eqref{bdrymetRindler}, we define
\begin{equation}
\tilde{\tau} \equiv \frac{\tilde{\beta}}{2\pi} \tau,\quad \tilde{x}_i \equiv \frac{\tilde{\beta}}{2\pi r_c}x_i \quad \implies \quad \left.ds^2\right|_{\partial\mathcal{M}} = \frac{4\pi^2 r_c^2}{\tilde{\beta}^2}\left(d\tilde{\tau}^2 + d\tilde{x}_i^2\right).
\end{equation}
This implies the following identification of each $L_i$ in terms of boundary data:
\begin{equation}
\tilde{L}_i = \frac{\tilde{\beta}}{2\pi r_c} L_i = \frac{\tilde{\beta} K}{2\pi}L_i \implies L_i = \frac{2\pi \tilde{L}_i}{\tilde{\beta}K}.
\end{equation}
There are additional bulk geometries consistent with \eqref{bdrymetRindler}. Besides the compactification of Rindler space, there are also $(d-1)$ vacuum geometries with toroidal boundary topology, each given by one of the spatial cycles capping off in the interior. For example, let us say $x_1$ caps off. Then the correct bulk geometry is given by
\be
ds^2 = d\t^2 + dr^2 + r^2 dx_1^2 + dx_a^2,\quad x_1 \sim x_1 + 2\pi,\quad x_a \sim x_a + L_a,\quad \t \sim \t + \b,
\ee
where we sum over $a = 2,\dots,d-1$. To match \eqref{bdrymetRindler} at $r = r_c$, we set
\begin{equation}
\tilde{\tau} \equiv \frac{\tilde{\beta}}{\beta}\tau,\quad \tilde{x}_1 \equiv \frac{\tilde{\beta} r_c}{\beta}x_1,\quad \tilde{x}_a = \frac{\tilde{\beta}}{\beta}x_a \longrightarrow \left.ds^2\right|_{\partial\mathcal{M}} = \frac{\beta^2}{\tilde{\beta}^2}\left(d\tilde{\tau}^2 + d\tilde{x}_1^2 + d\tilde{x}_a^2\right).
\end{equation}
Noting that $r_c = 1/K$, the bulk lengths $\beta$ and $L_a$ are fixed as
\begin{equation}
\beta = \frac{2\pi\tilde{\beta}}{\tilde{L}_1 K},\quad L_a = \frac{2\pi \tilde{L}_a}{\tilde{L}_1 K}.
\end{equation}
Observe that these vacuum geometries and the black hole geometry are all the same. They are distinguished by which cycle caps off in the interior. This is precisely what happens in AdS/CFT when the boundary theory is a $(d-1)$-dimensional torus: besides the compactified black brane solution there are $(d-1)$ vacuum geometries, each given by the AdS soliton \cite{Horowitz:1998ha} and distinguished by which spatial direction caps off in the interior. 

We will assume this exhausts the set of solutions which contribute to our boundary value problem. (The geometry known as ``hot flat space" $ds^2 = d\t^2 + dr^2 + dx_i^2$ with $\t \sim \t + \b$ can only accommodate $K = 0$ on constant radial slices and so does not appear in our analysis for general $K$.) To see which phases dominate as we vary our boundary data, we need to calculate the on-shell actions of the various solutions. For ease of notation, let us denote $\tilde{\tau}$ as $\tilde{x}_0$ and $\tilde{\beta}$ as $\tilde{L}_0$. Denoting the on-shell action of the geometry with a contractible $x_\alpha$-cycle by $I_\alpha$ (where $\alpha = 0,1,\dots,d-1$), we find that
\begin{equation}
I_\alpha = -c_0 \frac{\tilde{\beta} \tilde{L}_1 \cdots \tilde{L}_{d-1}}{\tilde{L}_\alpha^{d}} \equiv \tilde{L}_\alpha F_\alpha,
\end{equation}
where $F_\alpha$ is a high-temperature free energy with $\tilde{x}_\alpha$ taken as the thermal circle. Thus, the partition function takes the form written in \cite{Belin:2016yll} for the AdS/CFT problem, but in terms of the conformal length scales:
\begin{equation}
Z[\mathbb{T}^d] \approx \sum_{\alpha = 0}^{d-1} e^{-\tilde{L}_\alpha F_\alpha}.
\end{equation}
The sum is over the saddle points of the boundary value problem. This of course does not privilege any particular cycle as being the thermal circle. Nonetheless, we can return to using $\tilde{L}_0$ as the length of the thermal cycle $\tilde{\beta}$ and take note of a transition when $\tilde{\beta} = \text{min}\,\tilde{L}_i$ (i.e. the length of the smallest spatial cycle). Notice this exhibits a sort of modular invariance described by the group SL$(d,\mathbb{Z})$. This is essentially what happens in a $d$-dimensional CFT on the torus \cite{Belin:2016yll}, and it also enforces a Hawking--Page transition between the compactification of Rindler space and the true vacuum.

Still defining $\tilde{L}_0 = \tilde{\b}$ for clarity, and assuming $\tilde{\b} < \tilde{L}_i$ for all $i$, we can calculate the on-shell action of the path integral as
\begin{equation}
I_{\text{Rindler}} = -\frac{1}{8d\pi G}\int_{\partial\mathcal{M}} \sqrt{h}\,K = -\frac{(2\pi)^{d-1}}{4dG K^{d-1}} \frac{{\rm Vol}[\mathbb{T}^{d-1}]}{\tilde{\beta}^{d-1}},\quad {\rm Vol}[\mathbb{T}^{d-1}] \equiv \tilde{L}_1 \cdots \tilde{L}_{d-1}.\label{actplanarFlat}
\end{equation}
If we equate ${\rm Vol}[\mathbb{T}^{d-1}]$ to the volume of the unit $(d-1)$-sphere ${\rm Vol}[S^{d-1}]$, then this precisely agrees with the high-temperature limit of the spherical black holes. In particular, we have the same $c_0$ as obtained from the spherical geometry, \eqref{c0c1}. This result is expected. Since the high-temperature limit zooms us into the horizon of the black hole, we should be able to approximate the spherical horizon with a Rindler horizon, which is precisely what is happening here.

Interestingly, this solution also works in 3d flat space. From \eqref{actplanarFlat}, we get $c_0 = \pi/(4GK)$. This $c_0$ can be compared to the usual AdS$_3$ (with an infinite Dirichlet boundary) answer $c_0 = \pi\ell_{\text{AdS}}/(4G)$ \cite{allameh}. There is an SL$(2,\mathbb{Z})$ invariance in this case, which can be defined through its action on a modular parameter $\tau$. In our case of a rectangular torus this reduces to $\tilde{\b}/\tilde{L}$. 

The case where the boundary topology is a totally noncompact $\mathbb{R}^{d-1}$ gives pure Rindler space as our proposed dominant phase of the path integral. This is diffeomorphic to flat space. This case should be compared to that of the BTZ black string in AdS$_3$. This is the relevant geometry for the dual conformal field theory at finite temperature and infinite size, even though it is globally diffeomorphic to pure AdS$_3$. In higher-dimensional AdS/CFT a similar situation occurs when considering the CFT on hyperbolic space at inverse temperature $\b = 2\pi$. In the bulk the geometry is given by AdS-Rindler space, which is globally diffeomorphic to pure AdS spacetime.

\subsection{Solutions with hyperbolic horizons}\label{hypsection}
Now we consider our theory on a spatial hyperboloid, i.e.~we consider boundary conditions given by fixed $K$ and metrics conformal to $S^1 \times \mathbb{H}^{d-1}$. While there are generically no asymptotically flat geometries with hyperbolic horizons \cite{Hawking:1971vc}, we will find that our finite boundary conditions again give us a way out. To construct a solution to this problem, we begin with the hyperbolic black brane in AdS:
\be
ds^2 = \left(-1-\frac{\mu}{r^{d-2}}+\f{r^2}{\ell^2}\right) d\t^2 + \left(-1 - \frac{\mu}{r^{d-2}}+\f{r^2}{\ell^2}\right)^{-1} dr^2 + r^2 d\mathbb{H}_{d-1}^2,
\ee
where $d\mathbb{H}_{d-1}^2$ is the line element of hyperbolic space $\mathbb{H}^{d-1}$, and $\mu$ is the mass parameter. For $-\f{2}{d-2}\left(\f{d-2}{d}\right)^{d/2}\ell^{d-2} \leq \m < 0$, this geometry has two horizons, with a Penrose diagram presented on the left of Figure \ref{figs:hypPenrose}, while for $\m \geq 0$ it has just one. We now send the AdS scale $\ell \rightarrow \infty$ and get the metric
\be\label{inithype}
ds^2 = \left(-1-\frac{\mu}{r^{d-2}}\right) d\t^2 + \left(-1 - \frac{\mu}{r^{d-2}}\right)^{-1} dr^2 + r^2 d\mathbb{H}_{d-1}^2.
\ee
This metric has the wrong signature for $\mu \geq 0$ and no horizon. But notice that we can have a Euclidean metric with a horizon for $\m < 0$. This corresponds to taking the flat space limit of the negative-mass hyperbolic black brane in AdS. The radius of the outer horizon scales as $\ell$ in this limit and disappears, while the radius of the inner horizon scales as $(-\m)^\f{1}{d-2}$ and remains. The positive-mass black brane only has one horizon, whose location scales as $\ell$ and disappears in the flat space limit. It is therefore not a viable candidate for a flat space hyperbolic horizon. 

\begin{figure}[ht]
\centering
\begin{tikzpicture}[scale=1.35]
\draw[-] (-1,-1) to (-1,1);
\draw[-] (1,-1) to (1,1);
\draw[-,penrosered!80!black,dashed,very thick] (-1,1) to (0,0) to (-1,-1);
\draw[-,penrosered!80!black,dashed,very thick] (1,1) to (0,0) to (1,-1);
\draw[-,penroseblue!80!black,dashed,very thick] (-1,1) to (1,3);
\draw[-,penroseblue!80!black,dashed,very thick] (1,1) to (-1,3);
\draw[-,penroseblue!80!black,dashed,very thick] (-1,-1) to (1,-3);
\draw[-,penroseblue!80!black,dashed,very thick] (1,-1) to (-1,-3);
\draw[-,decorate,decoration={zigzag,amplitude=0.5mm,segment length=2.5mm}] (-1,1) to (-1,3);
\draw[-,decorate,decoration={zigzag,amplitude=0.5mm,segment length=2.5mm}] (1,3) to (1,1);
\draw[-,decorate,decoration={zigzag,amplitude=0.5mm,segment length=2.5mm}] (-1,-3) to (-1,-1);
\draw[-,decorate,decoration={zigzag,amplitude=0.5mm,segment length=2.5mm}] (1,-1) to (1,-3);
\node[penrosered!80!black] at (0.35,0.7) {\large$r_+$};
\node[penrosered!80!black] at (0.35,-0.7) {\large$r_+$};

\node[penroseblue!80!black] at (0.35,2+0.7) {\large$r_-$};
\node[penroseblue!80!black] at (0.35,2-0.7) {\large$r_-$};

\node[rotate=90] at (0,3.25) {\LARGE$\cdots$};
\node[rotate=90] at (0,-3.25) {\LARGE$\cdots$};

\draw[line width=0.55mm,->] (1.8,0) to (4.2,0);

\draw[draw=none,pattern = crosshatch,pattern color=penroseblue] (7,1) .. controls (6.45,0.25) and (6.45,-0.25) .. (7,-1) to (6,0) -- cycle;

\draw[-] (5,1) to (6,2) to (7,1);
\draw[-] (5,-1) to (6,-2) to (7,-1);
\draw[-,penroseblue!80!black,dashed,very thick] (5,1) to (7,-1);
\draw[-,penroseblue!80!black,dashed,very thick] (5,-1) to (7,1);
\draw[-,decorate,decoration={zigzag,amplitude=0.5mm,segment length=2.5mm}] (5,-1) to (5,1);
\draw[-,decorate,decoration={zigzag,amplitude=0.5mm,segment length=2.5mm}] (7,1) to (7,-1);
\draw[-] (7,1) .. controls (6.45,0.25) and (6.45,-0.25) .. (7,-1);
\node[penroseblue!80!black] at (6.35,0.7) {\large$r_h$};
\node[penroseblue!80!black] at (6.35,-0.7) {\large$r_h$};
\node at (6.775,0) {\large$r_c$};
\end{tikzpicture}
\caption{The Penrose diagram of the negative-mass hyperbolic black brane in AdS is presented on the left. A flat-space limit $\ell \rightarrow \infty$ gives the solution we use in this section, presented on the right with boundary cutoff $r_c$.}
\label{figs:hypPenrose}
\end{figure}

Altogether we consider the geometry
\be
ds^2 = f(r) d\t^2 + \frac{dr^2}{f(r)} + r^2 d\mathbb{H}_{d-1}^2,\qquad f(r) \equiv -1 - \frac{\mu}{r^{d-2}},\quad\, \m < 0\label{hypbh}
\ee
with $0 < r^{d-2} \leq - \mu$ to ensure that we have a Euclidean metric. Our geometry then keeps the region $r \in (r_c, r_h)$ where $r_h^{d-2} \equiv -\mu$. The singularity at $r = 0$ is excised, although the geometric description breaks down as $r_c \rightarrow 0$. The Penrose diagram of this geometry is presented on the right of Figure \ref{figs:hypPenrose}. 

Now with this geometry in hand, let us write the equations that determine $r_c$ and $r_h$ in terms of boundary data. At $r = r_c$, we have the following boundary metric and value of $K$:
\begin{equation}
ds^2 = f(r_c) d\tau^2 + r_c^2 d\mathbb{H}_{d-1}^2,\quad K = -\frac{d-1}{r_c \sqrt{f(r_c)}}\left[f(r_c) + \frac{r_c f'(r_c)}{2(d-1)}\right].
\end{equation}
Note the sign flip in $K$ relative to \eqref{indmetk}. This is because the normal vector at the boundary points in the $-\partial_r$ direction. We thus get the following equations for $\tilde{\beta}$ and $K$:
\be\label{hypbdryeqns}
\tilde{\b} = \f{4\pi r_h}{d-2} \frac{\sqrt{f(r_c)}}{r_c},\quad K = \frac{1}{2r_c \sqrt{f(r_c)}}\left[2(d-1) - d\left(\frac{r_h}{r_c}\right)^{d-2}\right].
\ee
So the formulas are basically the same as in the Schwarzschild geometry \eqref{sphereflateqns}. This time, however, $r_h > r_c$, which dramatically changes the solution space. The first difference is that we can find solutions for any $K \in \mathbb{R}$. For $K \neq 0$, we have either one solution or no solutions for each inverse conformal temperature $\tilde{\beta} > 0$, with $K > 0$ furnishing a maximal $\tilde{\beta}$ (i.e. a minimal temperature) and $K < 0$ furnishing a minimal $\tilde{\beta}$ (i.e. a maximal temperature). As long as we are below the maximal $\tilde{\b}$ for $K > 0$ and above the minimal $\tilde{\b}$ for $K < 0$, we have exactly one solution, and zero otherwise. The reason we do not have two solutions like in the spherical case is because $\tilde{\b}$ is monotonic in $r_c$ for $0 < r_c < r_h$. To contrast, in the spherical case with $r_c > r_h$ we have $\tilde{\b} = 0$ at $r_c = r_h$ and $r_c = \infty$, so the solution for $r_c$ in terms of $\tilde{\b}$ is double-valued. The fact that we have just one hyperbolic black hole given our boundary data mimics the situation in AdS/CFT with asymptotic Dirichlet boundary conditions.

We will also see below that for $K > 0$ we recover the same thermal effective action as in Section \ref{subsec:thermaleftschw}, but with sign flips at every odd power of curvature. For $K < 0$, since there is a minimal $\tilde{\beta}$, we cannot take a high-temperature limit where the thermal effective theory will be valid. We expect that the $K < 0$ solution is thermodynamically unstable, which can be explicitly verified through the exact solution in $d=3,4$ (the details of the $d=4$ case can be found in Appendix \ref{app:exact5d}). For $K = 0$, we will show that there are no solutions unless $\tilde{\beta}$ is finely tuned to a particular value, which gives an infinite family of degenerate solutions with zero free energy.

\subsubsection*{\textit{K} $\neq$ 0}

Depending on the sign of $K$, we have a particular allowed range of the ratio $\frac{r_h}{r_c} \in [1,\infty)$. Additionally, the inverse conformal temperature $\tilde{\beta}$ is monotonic in $\frac{r_h}{r_c}$, and so the sign of $K$ implies a particular range of the inverse conformal temperature $\tilde{\beta}$ as follows:
\begin{align}
K > 0 &&\implies&&& 1 < \frac{r_h}{r_c} < \left(2 - \frac{2}{d}\right)^{1/(d-2)} &&\implies&& \tilde{\b} < \frac{4\pi}{\sqrt{d(d-2)}} \left(2 - \frac{2}{d}\right)^{1/(d-2)},\\
K < 0 &&\implies &&&\frac{r_h}{r_c} > \left(2 - \frac{2}{d}\right)^{1/(d-2)} &&\implies&& \tilde{\b} > \frac{4\pi}{\sqrt{d(d-2)}} \left(2 - \frac{2}{d}\right)^{1/(d-2)}.
\end{align}
Thus, the $\tilde{\beta} \to 0$ limit needed for the thermal effective action is only consistent with $K > 0$. The geometric description breaks down as we approach the critical $\tilde{\b}$ separating $K > 0$ and $K < 0$ from either side, since $r_h$ and $r_c$ both go to zero in this limit, where the curvature is large.

The existence of a minimal $\tilde{\beta}$ for $K < 0$ can be argued a different way. First, note that the on-shell Euclidean action $I \sim -\int \sqrt{h} \,K$ implies that $K < 0$ gives a positive action. In a putative high-temperature limit, we would thus have a positive extensive term. The resulting high-temperature entropy $(\tilde{\b} \p_{\tilde{\b}} - 1)I$ would then be negative, contradicting the Bekenstein--Hawking area law.

Let's restrict to $K > 0$ to show that the thermal expansion is the same as that given by the large Schwarzschild black hole, but with relative sign flips at every term corresponding to an odd power of curvature. To see how this works, we first note that the equations for the hyperbolic black brane \eqref{hypbdryeqns} are related to those of the Schwarzschild black hole \eqref{sphereflateqns} by a phase, since their respective emblackening functions are the same up to an overall minus sign (once we recall that $\m > 0$ in the spherical case and $\m < 0$ in the hyperbolic case). More specifically, writing the boundary data of the hyperbolic geometry as $\{\tilde{\beta}_{\mathbb{H}},K_{\mathbb{H}}\}$ and that of Schwarzschild as $\{\tilde{\beta}_{\text{S}},K_{\text{S}}\}$, we have that they are related by an analytic continuation:
\begin{equation}
\tilde{\beta}_{\mathbb{H}} = i\tilde{\beta}_{\text{S}},\quad K_{\mathbb{H}} = -iK_{\text{S}}.\label{bdrydatarelation}
\end{equation}
Now, consider the action of the hyperbolic black brane:
\begin{equation}
I = -\frac{\tilde{\beta}_{\mathbb{H}} K_{\mathbb{H}}}{8d\pi G}\text{Vol}[\mathbb{H}^{d-1}]r_c^d.\label{tefthyp}
\end{equation}
We can replace the product $\tilde{\beta}_{\mathbb{H}} K_{\mathbb{H}}$ with $\tilde{\beta}_{\text{S}} K_{\text{S}}$. Meanwhile, assuming the expansion \eqref{expansions} for $r_c$ in the Schwarzschild geometry, we can use \eqref{bdrydatarelation} to write the series expansion of $r_c$ in the hyperbolic geometry in terms of the same coefficients.
\begin{equation}
r_c
= \frac{2\pi}{\tilde{\beta}_{\text{S}} K_{\text{S}}} \left(w_0 + w_1 \tilde{\beta}_{\text{S}}^2 + w_2 \tilde{\beta}_{\text{S}}^4 + \cdots\right)
= \frac{2\pi}{\tilde{\beta}_{\mathbb{H}} K_{\mathbb{H}}} \left(w_0 - w_1 \tilde{\beta}_{\mathbb{H}}^2 + w_2 \tilde{\beta}_{\mathbb{H}}^4 - \cdots\right).
\end{equation}
We see that we get minus signs on every odd $w$ coefficient, but the signs on every even $w$ coefficient stay the same. The magnitudes of the coefficients, however, stay the same. The high-temperature expansion of $r_c^d$ must have the same structure since $d$ is an integer.

Thus, the thermal effective action computed by \eqref{tefthyp} has, order-by-order, terms of the same magnitude as those of the expansion of the Schwarzschild action in Section \ref{subsec:thermaleftschw}. However, the hyperbolic action has a relative sign difference from the Schwarzschild action on every other term, and this directly corresponds to the fact that hyperbolic geometries have negative curvature rather than positive curvature. This result is entirely consistent with the prediction of thermal effective field theory. For the case $d=4$ where analytic solutions are straightforward see \eqref{tefthyperbolic}. 


So far, we have only discussed the black brane. But is there a thermal gas with hyperbolic boundary? Naively, we would write the metric for such a geometry by taking $\mu \to 0$ in \eqref{hypbh}. However, this clearly produces a semi-Riemannian geometry of signature $(2,d-1)$:
\begin{equation}\label{fakevac}
ds^2 = -d\tau^2 - dr^2 + r^2 d\mathbb{H}_{d-1}^2.
\end{equation}
So if the path integral only includes bulk Euclidean metrics (and not complex metrics), then the $\m \rightarrow 0$ limit of the geometries considered above does not provide a suitable vacuum solution. In terms of the ensemble, this means that there is not always a geometric dual to a state at fixed temperature. More specifically, for $K > 0$, the geometric description breaks down at low temperatures. 


Lastly, one may wonder what to make of the Lorentzian version of \eqref{hypbh}, and in particular its geodesic completion. Continuing the Lorentzian geometry past $r = r_h$ leads to a geometry whose causal structure is presented on the right of Figure \ref{figs:hypPenrose}. The geometry for $r \gg r_h$ is approximated as the analytic continuation $\t \rightarrow iz$, $r \rightarrow t$ of \eqref{fakevac}. This is a patch of Minkowski space $\mathbb{M}^{d+1}$ written as $\mathbb{R}\times \mathbb{M}^{d}$, with metric $ds^2 = dz^2 -dt^2 + t^2 d\mathbb{H}_{d-1}^2$.

\subsubsection*{\textit{K} = 0}

The case of $K = 0$ is somewhat singular, as it implies
\begin{equation}
\frac{r_h}{r_c} = \left(2 - \frac{2}{d}\right)^{1/(d-2)} \implies \tilde{\b} = \frac{4\pi}{\sqrt{d(d-2)}} \left(2 - \frac{2}{d}\right)^{1/(d-2)},
\end{equation}
where the inverse temperature is fixed by $r_h/r_c$. (The usual solution $r_c = \infty$ for $K = 0$ would result in a non-Euclidean metric.) Note that neither $r_c$ nor $r_h$ are independently fixed by the equations. This means that there is technically an infinity of solutions at a finely tuned $\tilde{\b}$. Furthermore, since both the bulk cosmological constant and $K$ vanish, the on-shell actions of all of these solutions vanishes, resulting in an infinite degeneracy. The interpretation of the $K=0$ case is unclear.

\section{Thermodynamic first law}\label{sec:thermo1st}

The thermal energy $\tilde{E} \equiv -\partial_{\tilde{\beta}} \log Z$ from Appendix \ref{appA} is given by 
\be
\tilde{E} =-\f{(d-1)\text{Vol}[\Sigma_{d-1}]r_c^d}{8\pi G}\left( \f{\sqrt{f(r_c)}}{r_c} - \f{K}{d}\right),\label{QEn}
\ee
where we recall that $\Sigma_{d-1}$ is the boundary spatial manifold. $\tilde{E}$ is related to the conserved energy as $\tilde{E} = r_c E^{\text{CBC}}$. The entropy is given by the usual Bekenstein--Hawking formula
\be\label{BHform}
S = (\tilde{\b}\partial_{\tilde{\b}}-1)I= \f{\text{Vol}[\Sigma_{d-1}]r_h^{d-1}}{4G}.
\ee
Using \eqref{QEn}--\eqref{BHform}, the total on-shell action evaluates to 
\be\label{grandcanans}
I = -\log Z = \tilde{\b}\tilde{E} - S.
\ee
We can also write this in terms of parameters normalized according to Schwarzschild time $\t$, namely $\b = \tilde{\b} r_c$, $E^{\text{CBC}} = \tilde{E}/r_c$, for which we have the more standard
\be
I = -\log Z = \b E^{\text{CBC}} - S.
\ee
With conformal boundary conditions the potential $\b$ is not held fixed, so \eqref{grandcanans} is a more natural way to write the answer.

We can also derive a thermodynamic first law:
\be\label{firstlawcbc}
d\tilde{E} = \tilde{T} dS + \tilde{V} dK.
\ee
A first law of this form was established for conformal boundary conditions in de Sitter space in \cite{Anninos:2024wpy}. If we think of $K$ as something akin to a pressure (more on this below), then \eqref{firstlawcbc} indicates that $\tilde{E}$ is identified with something like the enthalpy\footnote{The identification of the conserved energy with enthalpy also occurs in ``extended black hole thermodynamics," where the cosmological constant is treated as a dynamical variable proportional to the pressure $p$ of the system (see \cite{Kubiznak:2016qmn} for a review).} of the system and can be written in terms of the so-called internal energy $\tilde{U}$ as
\begin{equation}
\tilde{E} = \tilde{U} + K \tilde{V}.
\end{equation}
Mathematically, the $K\tilde{V}$ term in $\tilde{E}$ is equal to the second term in \eqref{Enformal} multiplied by $r_c$. Thus, $\tilde{U}$ equals the energy with Dirichlet boundary conditions, multiplied by $r_c$.

We stress that the form \eqref{firstlawcbc} of the first law is not commensurate with our boundary conditions, which keep $K$ fixed but not $\tilde{V}$. For a first law with $K$ held fixed, we switch to the internal energy:
\be\label{firstlawdbc1}
d\tilde{U} = \tilde{T} dS - K d\tilde{V}.
\ee
In other words, we have the thermodynamic relations
\be
\left(\f{\p \tilde{U}}{\p S}\right)_{\tilde{V}} = \tilde{\b}^{-1},\quad \left(\f{\p \tilde{U}}{\p \tilde{V}}\right)_{S} = -K.
\ee
This can be verified, for example in the spherical case $\Sigma_{d-1} = S^{d-1}$, using \eqref{sphereflateqns} and \eqref{BHform} with the function $f(r) = 1-r_h^{d-2}/r^{d-2}$ \eqref{bhflat}. The volume comes out to be, as expected, proportional to $r_c$ times the spatial volume of the boundary:
\be
\tilde{V} = \f{(d-1)\text{Vol}[\Sigma_{d-1}]}{8\pi G d}\, r_c^d.
\ee
The extra factor of $r_c$ compared to a spatial volume is the usual one to switch to tilded quantities, and this factor gives a scaling appropriate for a spatial volume in the bulk.

\subsubsection*{Comparing $\boldsymbol{K}$ to pressure}

While the $K d\tilde{V}$ term in the first law \eqref{firstlawdbc1} looks like a $pdV$ term that might appear in black hole thermodynamics, they are not exactly the same because the length-scaling of the volumes differ, as we now discuss.

The following is how one would get a standard $pdV$ term in black hole thermodynamics (with fixed cosmological constant). For clarity, we will consider all thermodynamic variables as referencing a putative boundary field theory. In particular, the spatial volume $V$ will refer to a $(d-1)$-dimensional boundary volume. To contrast, the volume $\tilde{V}$ has the units of a $d$-dimensional spatial volume in the bulk (divided by $G$). We will work at finite volume and choose Dirichlet boundary conditions on the metric. To remind ourselves in field theory language why we expect $p$ to appear with the spatial volume $V$, we calculate the source term 
\be
 \int d^d x \sqrt{h} \,\eta_{ij} T^{ij} \propto \int d^d x \sqrt{h} \,T^i_i = \b V T^0_0 + \b V T^a_a = \b\left[E + (d-1)\, pV\right].\label{sourcepv}
\ee
where $\eta_{ij}$ is a source that is chosen to be isotropic with respect to the background metric, $\eta_{ij} \propto h_{ij}$. Spherical symmetry implies that we have a single rotationally invariant pressure $p = T^a_a/(d-1)$, with the trace taken only over the spatial components of the boundary stress tensor. (Recall the diagonal components of the stress tensor correspond to the pressure.)

In \eqref{sourcepv}, we have the pressure appearing with the standard spatial volume $V$. When calculating $Z = \Tr[e^{-\beta H}]$, we only source $\eta_{00} T^{00}$, which gives the $\b E$ term when integrated. To check the explicit form of the first law, we can use the ordinary Brown--York stress tensor obtained from the action appropriate for Dirichlet boundary conditions,
\be
T_{\m\n} = -\f{1}{8 \pi G}(K_{\m\n} - K h_{\m\n}),
\ee
to obtain the following energy and pressure,
\begin{align}
E &= - \f{(d-1)\text{Vol}[\Sigma_{d-1}] r_c^{d-2}{\sqrt{f(r_c)}}}{8\pi G},\\
p &= \f{1}{d-1}\, T^a_a = -\f{1}{8(d-1)\pi G}[K^a_a - (d-1)K].
\end{align}
Finally we have the volume $V = \text{Vol}[S^{d-1}]r_c^{d-1}$. Together these quantities satisfy 
\be\label{firstlawDBC}
dE = TdS - (d-1)pdV.
\ee
So, we see that the pressure corresponds to the diagonal components of the (spatial) stress tensor, as is usually the case in quantum field theory. The volume in black hole thermodynamics simply corresponds to the volume of the finite-cutoff boundary. A first law of the form \eqref{firstlawDBC} appeared in \cite{York:1986it}, although the pressure was not understood as coming from $T^a_a$. 

Furthermore, we can define the enthalpy $H = E +(d-1)pV$ to get
\be
dH = T dS +(d-1)V dp.
\ee
Notice that we have $\tilde{U} = r_c E$ in terms of the Dirichlet energy $E$, so we have the closely related first law \eqref{firstlawdbc1}, which we can write as 
\be
d(r_c E) = r_c T\,dS - \f{d-1}{8 \pi G d}\, K d(r_c V).
\ee
We also note that we can get a $pdV$ term without even working at finite cutoff in AdS/CFT. The idea is to calculate the partition function for the boundary theory at finite size and finite temperature. Then, we are free to vary the size, and a $pdV$ term will result. For example, the thermal energy of the AdS black brane toroidally compactified with a $(d-1)$-dimensional spatial volume $V$ is $E = c_0(d-1)V/\b^{d-1}$, which gives $\p E/\p V = c_0(d-1)/\b^{d-1}$. This indeed corresponds to the pressure $p = T^a_a$ calculated from the Brown--York stress tensor.

The key point here as before is that a ``standard" volume term should be interpreted as a spatial volume in the dimensionality of the boundary, i.e. it scales as $r_c^{d-1}$. By contrast, $\tilde{V}$ appearing in the first law \eqref{firstlawdbc1} scales as $r_c^d$ (divided by $G$), i.e. it is like a bulk spatial volume.

\section{The sphere partition function}\label{sec:spherePart}

Another interesting quantity which probes the details of a putative boundary dual is the sphere partition function. For conformal field theories this is equivalent to the entanglement entropy in the vacuum state of flat space across a spherical entangling surface \cite{Casini:2011kv}, which is known to be monotonic under renormalization group flow for $d \leq 4$ \cite{Zamolodchikov:1986gt, Casini:2004bw, Komargodski:2011vj, Casini:2012ei, Casini:2017vbe}. In our context we fix boundary conditions in the bulk path integral such that $K$ is fixed and the conformal class of geometries has $S^d$ as a representative. We propose that this path integral is dominated by empty flat space:
\be
ds^2 = d\rho^2 + \rho^2 d\Omega_d^2.
\ee
Picking a cutoff $\rho_c$ gives
\be
K = \f{d}{\rho_c}\,
\ee
on the cutoff surface. Thus fixing the conformal class of metrics tells us that we are at some unknown but constant $\rho_c$, while fixing $K$ then fixes $\rho_c$. The bulk part of the on-shell action vanishes since $R = 0$, while the boundary part contributes
\be
\log Z[S^d] = \f{\a\,\text{Area}[S^{d}]}{8 \pi G}\, \rho_c^d\, \f{d}{\rho_c} = \f{ \text{Area}[S^{d}]}{8 \pi G} \rho_c^{d-1}= \f{ \text{Area}[S^{d}]d^{d-1}}{8 \pi G K^{d-1}} .
\ee
where we used $\a = 1/d$. For $d = 2$ this gives
\be
\log Z = \f{\rho_c}{2G} = \f{1}{GK}.
\ee
The form $\log Z = \rho_c/(2G)$ looks just like the answer in AdS$_3$ (with the usual asymptotic Dirichlet cutoff) when written in coordinates $d\rho^2 + \sinh^2 \rho \, d\Omega^2$. In that case, however, $\rho_c$ has a logarithmic relation to the radius of the sphere $\rho_c \sim \log R$. This is where the conformal anomaly comes from. In the flat space case $\rho_c$ is the radius of the sphere, so we do not get the $\log R$ scaling. Furthermore, the appropriate boundary variable to phrase the answer in is $K$. (See also \cite{Anninos:2024wpy} for the $S^2$ partition function in (A)dS$_3$ with conformal boundary conditions.)

This answer may look a bit troubling, since it seems inconsistent with $T^\m_\m = 0$. In other words, we expect 
\be
\f{d \log Z}{d \log R} \sim \int d^d x\,\sqrt{h} \,T^\m_\m.
\ee
However, notice that the tracelessness of the stress tensor depended on the fact that $K$ is kept constant in the variational problem. In this case, changing the radius of the sphere would change $K$; insisting that $K$ remain fixed means $\f{d\log Z}{d\log R} = 0$.

Not having a microscopic, boundary understanding of $K$ means this argument is a purely mathematical illustration of the consistency between the trace of $T$ and varying the conformal mode of the sphere. For example, for Dirichlet boundary conditions, we understand that we are allowed to change the boundary conditions when calculating $\f{d \log Z}{d\log R}$; that is what the derivative instructs us to do, and it changes the ``fixed" Dirichlet boundary condition. In this case, however, we interpret the derivative as requiring us to vary the intrinsic part of the boundary data---which is the conformal metric $h^{-1/d}h_{\m\n}$---while keeping fixed the extrinsic part $K$. Since the conformal metric is independent of the radius the derivative vanishes. 

\section{Boundary conditions and extensivity}\label{sec:bdryCond}
One of the remarkable features of conformal boundary conditions is that they lead to extensive scaling of the high-temperature entropy.
In this section we will provide an explanation for this phenomenon for a large class of horizons whose spacetimes can be put into the form \eqref{bhgeneral}. Interestingly, this explanation does not capture the extensive scaling in de Sitter \cite{Anninos:2024wpy} or anti-de Sitter space.

We will also examine the generalized one-parameter family of well-posed boundary conditions recently discussed by \cite{Liu:2024ymn}. This family reduces to conformal boundary conditions at a specific tuning. We find a high-temperature entropy that scales as $S \sim T^{\d-1}$, with $\d = d(1-2p)/(1-2pd)$ and $p>0$ parameterizing the family of boundary conditions. As noted in \cite{Liu:2024ymn}, one finds a negative specific heat for $p > 1/(2d)$. For $0< p < 1/(2d)$ the entropy is superextensive, scaling with an effective dimensionality $\d-1 > d-1$. This is analogous to systems that violate hyperscaling. $p = 0$ corresponds to conformal boundary conditions, and as expected taking this value yields extensive scaling in the entropy.

\subsection{Universality of extensivity with conformal boundary conditions}
The extensivity of the high-temperature entropy with conformal boundary conditions is a somewhat general phenomenon. We will show that it is always true for horizons satisfying $\b \sim r_h^k$ for $k > -1$ (see \eqref{bhgeneral} below). The reason it occurs is because the high-temperature limit is like a near-horizon limit in which the geometry can be approximated by Rindler space. As shown in \eqref{actplanarFlat}, the entropy in Rindler space (for conformal boundary conditions) is extensive. This argument requires a bit of care, since the high-temperature limit is not a standard $r_h \rightarrow r_c$ limit at fixed $r_c$, as this would force $K \rightarrow \infty$. Instead, we need to take both $r_h$ and $r_c$ to infinity as $r_h \rightarrow r_c$. To see that the Rindler approximation is still valid in this particular limit, we consider the following class of geometries: 
\begin{equation}\label{bhgeneral}
ds^2 = -f(r) dt^2 + \frac{dr^2}{f(r)} + r^2 d\Sigma_{h}^2,\qquad t \sim t + i \b.
\end{equation}
$\Sigma_h$ is the geometry of the horizon, and $f$ is a function with simple root $r_h$ that sets the size of the horizon. Now we define $y \equiv \sqrt{\b/\pi}\sqrt{r - r_h}$, where $\beta = 4\pi/f'(r_h)$. The horizon is at $y =0$, and the near-horizon regime is $y^2\ll r_h \b$. In this approximation, we find the universality of the near-horizon Rindler geometry for nonextremal horizons: 
\be
ds^2 \approx -\f{4\pi^2}{\b^2} y^2 dt^2 + dy^2 + r_h^2 d\Sigma_h^2 = -y^2 dT^2 + dy^2 + r_h^2 d\Sigma_h^2,\qquad (y^2 \ll r_h \b ).\label{approxRindler}
\ee
The time coordinate $t$ is the same one that appears in the original black hole metric, and $t \sim t + i\b$. One can also define a coordinate $T \sim T + 2\pi i$ as done above. Importantly, the geometry is a direct product between 2d Rindler 
space and a $(d-1)$-dimensional Euclidean space with the geometry of the horizon.

Now, we use the solution \eqref{approxRindler} to solve the boundary value problem with conformal boundary conditions. We write
\be
\tilde{\beta} = \f{2\pi y_c}{r_h},\quad K = \f{1}{y_c}.\label{bdryrindler0}
\ee
We want to take $\tilde{\b}\rightarrow 0$ at fixed $K$, which according to the equations above means we take $r_h \rightarrow \infty$ at fixed $y_c$ (these length scales can be compared to a fixed $\ell_{Pl}$). This respects the approximation $y^2 \ll r_h\b$ needed for \eqref{approxRindler} as long as $y_c^2 \ll r_h \b$. So we need
\be \label{yc-limits}
y_c \ll r_h \qquad {\rm and } \qquad y_c \ll \sqrt{r_h \b},
\ee
which can be satisfied as long as $\b \gtrsim r_h^k$ for $k > -1$. Of course, if we are willing to take $K \rightarrow \infty$ via $y_c \rightarrow 0$, then we can ensure we reach a Rindler regime; this is simply the ordinary Rindler limit where we take $r_h \rightarrow r_c$ at fixed $r_c$. 

We can see the extensivity immediately from \eqref{bdryrindler0}: $S \sim r_h^{d-1} \sim \tilde{\b}^{1-d}$. Choosing conformal boundary conditions was essential here, as it brought in the factor of $r_h$ in $\tilde{\b}$. We can also compute the on-shell Euclidean action of the near-horizon Rindler geometry. The result is
\begin{equation}
I = -\frac{2\pi K y_c r_h^{d-1}}{8d\pi G}\,{\rm Vol}[\Sigma_h] = -\int_{\Sigma_h} \sqrt{g}\,\frac{c_0}{\tilde{\beta}^{d-1}},\quad c_0 = \frac{(2\pi)^{d-1}}{4dG K^{d-1}},
\end{equation}
i.e. extensive scaling, regardless of the horizon shape. We expect the saddle we used to be the dominant high-temperature solution in the path integral.

The case of anti-de Sitter space with conformal boundary conditions is different. In particular, $\b \sim r_h^{-1}$ in the high-temperature limit, so \eqref{yc-limits} cannot be satisfied. What happens instead is that the high-temperature limit pushes us into a black brane regime where we recover extensivity from the thermodynamic properties of AdS black branes.\footnote{Due to the choice of conformal boundary conditions, one has to check carefully that the high-temperature limit still pushes one into the AdS black brane regime, in the same sense that we had to check that the high-temperature limit above pushed us into the Rindler regime.} The case of de Sitter space is similar---a high-temperature limit can be taken by utilizing the negative-mass Schwarzschild-de Sitter black holes with the boundary excising the singular region, and this gives (up to sign flips in $g_{tt}$ and $g_{rr}$) the same metric as the AdS black brane. These cases will be elaborated upon in future work. 

\subsection{Boundary conditions with superextensive scaling}

Recently, \cite{Liu:2024ymn} has proposed a one-parameter family of boundary conditions that are elliptic, meaning that they are well-posed in the same sense as conformal boundary conditions.\footnote{Regarding notation, we stress that our $d$ is the number of boundary dimensions, while the $d$ in \cite{Liu:2024ymn} is the number of bulk dimensions.} The boundary value problem is
\begin{equation}
G_{\mu\nu} = 0,\quad \delta(h^{-1/d}h_{\mu\nu}) = 0,\quad \delta(h^{p} K) = 0.\label{bcproblemsantos}
\end{equation}
For convenience, let us define $K_p \equiv h^p K$. The point $p = 0$ corresponds to the conformal boundary condition, whereas taking $p \to \infty$ corresponds to Dirichlet boundary conditions. Furthermore, the action consistent with this boundary condition is
\begin{equation}
I = -\frac{1}{16\pi G}\int_{\mathcal{M}} \sqrt{g}\,R - \frac{\alpha_p}{8\pi G}\int_{\partial\mathcal{M}}\sqrt{h}\,K,\qquad \alpha_p \equiv \frac{1-2pd}{d(1-2p)}.\label{actgenp}
\end{equation}
We will see that finite $p \neq 0$ gives a thermal effective action that does not scale extensively. Rather, it scales superextensively for $0 < p < 1/(2d)$---the values of $p$ for which specific heat is positive \cite{Liu:2024ymn}---indicating a violation of hyperscaling \cite{Fisher:1986zz}. In terms of the traditional critical exponent of hyperscaling violation $\q$, which parameterizes the high-temperature entropy $S \sim T^{d-\q-1}$, we have
\begin{equation}
\q = \frac{2d(d-1)p}{2dp-1} < 0.
\end{equation}
While this is not a completely exotic phenomenon in gauge-gravity duality \cite{Ogawa:2011bz,Huijse:2011ef}, it is interesting to note that the scaling is extensive precisely when $p = 0$, which corresponds to the conformal boundary condition.

We now exhibit this superextensivity explicitly. In the boundary value problem \eqref{bcproblemsantos}, we fix a conformal class of geometries $\l^2(d\tilde{\tau}^2 + d\Omega_{d-1}^2)$ with $\tilde{\tau} \sim \tilde{\tau} + \tilde{\beta}$. At high temperature, the dominant solution is proposed to be the Schwarzschild metric in the bulk \cite{Liu:2024ymn}, which we repeat for convenience:
\begin{equation}
ds^2 = f(r)d\tau^2 + \frac{dr^2}{f(r)} + r^2 d\Omega_{d-1}^2,\quad f(r) \equiv 1 - \frac{r_h^{d-2}}{r^{d-2}}.
\end{equation}
Just as for conformal boundary conditions, the boundary is at $r = r_c$, with $\l = r_c$. The equations that determine the radii $r_h$ and $r_c$ are
\begin{equation}
\tilde{\beta} = \frac{4\pi r_h}{d-2} \frac{\sqrt{f(r_c)}}{r_c},\qquad K_p = \frac{1}{2r_c^{1-2pd} \sqrt{f(r_c)}}\left[2(d-1) - d\left(\frac{r_h}{r_c}\right)^{d-2}\right].\label{constraintsp}
\end{equation}
The equation for the inverse conformal temperature is the same as before, but the second formula is different because we are fixing $h^{p} K$. The extra factor of $r_c^{2pd}$ comes from the fact that the determinant of the boundary metric is $h = \l^{2d} = r_c^{2d}$ in the classical solution.

Now, let us compute $r_h$ and $r_c$ in a high-temperature expansion. There are still two black holes, but we focus on the large one for which $r_h/r_c \to 1$ in the high-temperature limit. The first step is to solve for the ratio $r_h/r_c$ in a series expansion using the first equation of \eqref{constraintsp}:
\begin{equation}
\frac{r_h}{r_c} = 1 - \frac{(d-2)}{16\pi^2}\tilde{\beta}^2 - \frac{(d+1)(d-2)^2}{512\pi^4}\tilde{\beta}^4 + \cdots.
\end{equation}
We can then use the second equation to solve for $r_c$, then multiply the answer against the above series to get $r_h$. This procedure yields
\begin{equation}
\begin{split}
r_h
&= \left(\frac{2\pi}{\tilde{\beta}K_p}\right)^{1/(1-2pd)}\left[1 + \frac{(d-2)(d-2+2pd)}{16\pi^2(1-2pd)}\tilde{\beta}^2 + \cdots\right],\\
r_c
&= \left(\frac{2\pi}{\tilde{\beta}K_p}\right)^{1/(1-2pd)}\left[1 + \frac{(d-1)(d-2)}{16\pi^2(1-2pd)}\tilde{\beta}^2 + \cdots\right].
\end{split}
\end{equation}
With these expressions, we can compute the on-shell action of the large black hole by plugging into \eqref{actgenp}. We find that
\begin{align}
I
&= -\frac{\alpha_p K_p}{8\pi G} \tilde{\beta}r_c^{d(1-2p)} \text{Vol}[S^{d-1}]\nonumber\\
&= -\frac{\text{Vol}[S^{d-1}]}{4G}\left(\frac{2\pi}{\tilde{\beta} K_p}\right)^{(d-1)/(1-2pd)}\left[\frac{1-2pd}{d(1-2p)} + \frac{(d-1)(d-2)}{16\pi^2}\tilde{\beta}^2 + \cdots\right],
\end{align}
where $\text{Vol}[S^{d-1}]$ is the volume of the unit $(d-1)$-sphere. Assuming that this dominates the canonical ensemble at high temperatures \cite{Liu:2024ymn}, we can write the answer in terms of a slightly modified thermal effective action:
\begin{equation}
I = \int_{\Sigma} d^{d-1}x\,\sqrt{g}\left[-\frac{c_0}{\tilde{\beta}^{\delta-1}} + \frac{c_1}{\tilde{\beta}^{\delta - 3}} R + \cdots\right],
\end{equation}
where $\delta \equiv d - \theta$ is the ``effective" dimension. This and the first two Wilson coefficients are
\begin{equation}
\delta = \frac{d(1-2p)}{1-2pd},\quad c_0 = \frac{(2\pi)^{\delta - 1}}{4\delta GK_p^{\delta - 1}},\quad c_1 = -\frac{(2\pi)^{\delta - 3}}{16 G K_p^{\delta - 1}}.
\end{equation}
Interestingly, these Wilson coefficients are ``morally" the same as before \eqref{c0c1}. The main differences are that we use $K_p$ instead of $K$ and $\delta$ instead of $d$.

\section{Conclusions}\label{conclusions}

The primary aim of our analysis was to learn more about the putative boundary dual of flat-space gravity with conformal boundary conditions. We did this by comparing to the structure of the thermal effective action for the boundary theory place on a spatial sphere, hyperbolic space, and flat space. For the case of the sphere, we were able to use the ordinary Schwarzschild solution and found a series of subextensive terms in the free energy separated by powers of $\tilde{\b}^2$, exactly in line with the effective action. For the boundary theory on hyperbolic space, we used a bulk solution that could be understood as the flat-space limit of the negative-mass hyperbolic black brane in AdS. This solution gave the same free energy, except with sign flips at every other term in a small-$\tilde{\b}$ expansion. These sign flips are exactly predicted from the thermal effective action, as they correspond to background terms with an odd number of curvatures. For the boundary theory on flat space, our bulk solution was given by Rindler space. This gave the leading extensive piece in the free energy, again as predicted by the effective action. While we saw agreement with the structure of the effective action in these cases, we also found that the conjectured QFT bound $c_1 \geq 0$ is violated. 

For $d=3$, this violation can potentially be explained by the presence of gapless modes. Gapless modes would contribute to the $\tilde{\b}^0$ piece of the effective action, which at $d = 3$ is precisely the same order as $c_1$. In fact, this is what happens in the 3d free scalar field theory---the $\b^0$ piece of the effective action is negative, but that negativity comes entirely from the gapless sector, and we actually have $c_1 = 0$. The coefficient of the full subextensive term is only equal to $c_1$ when $\igapless = 0$. However, this explanation would not work for $d > 3$; there, any gapless sector would continue to contribute at order $\tilde{\b}^0$, which is not the leading subextensive piece. 

What we propose instead is that this is a sharp signal that the boundary theory is not an ordinary QFT. Instead, it is a QFT coupled to a metric degree of freedom. This is expected from a couple of different perspectives: (1) since we do not fix the conformal mode of the boundary metric, we expect to have to integrate over it. This therefore couples the boundary QFT to a metric degree of freedom, the Weyl mode; (2) gravity is generically not expected to decouple at a finite-cutoff boundary; and (3) the conjectured QFT bound $c_1 \geq 0$ implies an Einstein--Hilbert term of the \emph{opposite} sign compared to Einstein gravity. While this is no problem for an effective action of background terms, a dynamical theory of gravity with such a sign would be unstable. (Notice though that this is the Einstein--Hilbert term of the dimensionally reduced effective theory.)

This begs the question of how general the result $c_1 < 0$ is. A natural strategy to disprove our interpretation is to consider conformal boundary conditions in anti-de Sitter spacetime. We know that $c_1 \geq 0$ in ordinary AdS/CFT \cite{allameh}, namely with Dirichlet conditions at the asymptotic boundary. Since conformal boundary conditions let us vary the cutoff from infinity (where $c_1 \geq 0$) to a finite cutoff (where we expect $c_1 < 0$ by the intuition above), we should see $c_1$ change sign at some point. This would imply there is some range where we have a finite-cutoff theory and $c_1 >0$, violating our intuition. In fact, what happens instead is that the high-temperature limit $\tilde{\b} \rightarrow 0$ (needed to define $c_1$) and the $K \rightarrow d$ limit (needed to push the finite-cutoff to the asymptotic AdS boundary) do not commute. Furthermore, for any $K > d$, the high-temperature limit leads to $c_1 < 0$, consistent with our interpretation above. Setting $\ell_{\text{AdS}} = 1$, the precise values for the first two Wilson coefficients are
\begin{align}
c_0 &= \f{1}{4dG}\left[\f{4\pi}{d^2}\left(K - \sqrt{K^2 - d^2}\right)\right]^{d-1},\label{c0ads}\\
c_1 &= \frac{1}{4d(d-2)G}\left[\frac{4\pi}{d^2}\left(K - \sqrt{K^2 - d^2}\right)\right]^{d-3} \left[1 - \left(\frac{K + \sqrt{K^2 - d^2}}{2\sqrt{K^2 - d^2}}\right)^{\f{d-2}{d}}\right].
\end{align}
These can be obtained from the AdS-Schwarzschild black hole and an ansatz for the high-temperature solution analogous to \eqref{expansions}. Notice that as $K \to d$, the $c_1$ above diverges as $1/(K-d)^{(d-2)/(2d)}$. By contrast, fixing $K = d$ first recovers the infinite cutoff, after which taking a high-temperature limit gives the finite AdS/CFT answer $c_1 = \frac{1}{4d(d-2)G} \left(\frac{4\pi}{d}\right)^{d-3} > 0$.

The Wilson coefficients $c_0$ and $c_1$ for flat space and de Sitter space can be obtained from the expressions above. To see this we first restore the AdS scale as $K \rightarrow K \ell_{\text{AdS}}$, $G \rightarrow G/\ell_{\text{AdS}}^{d-1}$. To recover the flat space answer we take $\ell_{\text{AdS}} \rightarrow \infty$, which gives agreement with our expressions \eqref{c0c1}. The answers for de Sitter space can be obtained by an analytic continuation $\ell_{\text{AdS}}\rightarrow i\ell_{dS}$. This again gives a value of $c_1$ satisfying $c_1 < 0$, now for arbitrary real $K$ (i.e. including negative values). The value for $c_0$ in $d=3, 4$ was previously obtained in \cite{Anninos:2024wpy}. The dS and AdS results will be elaborated upon elsewhere.

It is important to remember that the high-temperature limit in our setup $\tilde{\b} \rightarrow 0$ corresponds to a limit where the cutoff surface goes to infinity. This means that, for example in the AdS context where this also happens, the consistency with the thermal effective action only suggests locality of the theory in this limit. The generic situation of finite temperature corresponds to finite cutoff, where the expectation is that the boundary dual is nonlocal. 

\subsection*{Boundary counterterms}

One potential ambiguity worth mentioning is that of boundary terms that can be added beyond the extrinsic curvature term needed for the variational principle. In flat space, one usually adds a reference extrinsic curvature term to the Gibbons--Hawking--York term to cancel infinite-volume divergences, whereas in the context of AdS/CFT, one instead adds so-called holographic counterterms \cite{Emparan:1999pm}. However, in our calculations, for fixed $K \neq 0$ and finite $\tilde{\b}$ we have a finite cutoff and therefore no divergences. As we take $\tilde{\b} \rightarrow 0$, the cutoff goes to infinity and divergences appear. But the only divergences are the expected high-temperature ones. In particular, the leading divergence is the extensive term scaling as $1/\tilde{\b}^{d-1}$. Furthermore, simple intrinsic counterterms like a cosmological constant or Ricci scalar $R$ lead to a modified variational principle and so cannot be added when considering conformal boundary conditions.

The fact that we do not add any boundary terms beyond the one needed for a consistent variational principle means that our formula for the energy \eqref{E-CBC} does not match the usual ADM energy in the limit $r_c \rightarrow \infty$ at fixed $r_h$. In fact, it goes to negative infinity in this limit. Interestingly, using \eqref{E-CBC}, we can see that the energy turns negative precisely at $r_c f'(r_c)/[2 f(r_c)] = 1$, which corresponds to $z = r_h/r_c = (2/d)^{1/(d-2)}$. This is the same value of $z$ for which our large and small black holes are degenerate. In other words, the black holes with positive energy have positive specific heat, and the ones with negative energy have negative specific heat.

\subsection*{Grand canonical ensemble}

Another interesting direction is to add an electric potential to the ensemble \cite{upcoming}. The bulk theory in this case needs to be upgraded to Einstein--Maxwell theory, and the relevant solutions will be patches of the Reissner--Nordstr\"om black hole. A novel feature in this case is that the patches can correspond to regions $r_c < r < r_-$ for $r_-$ the radius of the inner horizon (in addition to the usual $r_+ < r < r_c$ for $r_+$ the radius of the outer horizon). These still yield Euclidean geometries that contribute to the path integral. In fact, these patches are akin to our hyperbolic solutions, which also covered a range $r_c < r < r_h$. These solutions are more appropriately thought of as cosmic horizons instead of black hole horizons, because the horizon is a maximal surface instead of a minimal surface. This problem also brings in the Maxwell field, for which appropriate boundary conditions must be chosen that play well with conformal boundary conditions on the metric. 

We can also try to check the predictions of the thermal effective action at finite angular potential. This requires black hole solutions with rotation. To capture the leading correction from rotation, it suffices to investigate the regime of small angular velocity. The naive guess is to use the Kerr solution, since the AdS-Kerr solution was used to extract 
$c_2$ (the coefficient of the Kaluza--Klein gauge field strength term) in the context of AdS/CFT \cite{Benjamin:2023qsc,allameh}. The problem with this approach is that we cannot generically fit this solution into our boundary value problem. Constant radial slices in Boyer--Lindquist coordinates do not give a geometry conformal to $S^1 \times S^{d-1}$, nor do they have constant extrinsic curvature. We can pick a nontrivial hypersurface where the radius depends on the angular coordinates to either accommodate constant extrinsic curvature, or constant intrinsic curvature, but not both.

The correct solution for this problem has already been constructed numerically \cite{Adam:2011dn}. These solutions are not Kerr black holes; since they are solutions inside of a finite cavity, they can circumvent the uniqueness theorem regarding Kerr black holes by not being asymptotically flat. It would be interesting to solve the equations analytically in a small angular velocity expansion to extract $c_2$, which may also have a sign constraint in QFT \cite{Benjamin:2023qsc, allameh}. Note however that, unlike with spherical symmetry, the rotation parameter of a Wick-rotated axisymmetric solution will need to be complexified to ensure reality of the metric.

\subsection*{Boundary interpretation of $K$}

An outstanding question is an explanation of the meaning of $K$ from the boundary theory. In many bulk expressions, e.g. \eqref{genericActExp}--\eqref{c0c1}, $K$ appears with $G$ in a way reminiscent to $\ell_{\text{AdS}}$. This suggests that $K$ (or more precisely $1/(GK^{d-1})$) may be related to a count of the number of degrees of freedom (although this interpretation is muddied when other scales appear, e.g.~$\ell_{\text{AdS}}$ in \eqref{c0ads}). Another perspective that supports this interpretation is that $K$ is conjugate to the volume density $\sqrt{h}$, as can be seen in \eqref{Ivary} or in the thermodynamic context in \eqref{firstlawcbc}--\eqref{firstlawdbc1}. There is precedent for interpreting $K$ as a count of degrees of freedom, as precisely this was proposed \cite{Sahakian:1999bd} around discussions of the holographic $c$-theorem in AdS/CFT \cite{Girardello:1998pd, Freedman:1999gp}, with the monotonicity of $K$ assured by Raychaudhuri's equation. Indeed, for the traditional domain wall flows considered in the holographic $c$-theorem,
\be
ds^2 = dr^2 + e^{2A(r)} dx_\m^2,
\ee
we have $K = A'(r)$, with $\p_r K = A''(r) < 0$ guaranteed by the null energy condition. 

The interpretation of $K$ in terms of a count of degrees of freedom makes negative values of $K$ particularly puzzling. It is interesting that such cases seem to appear hand-in-hand with cosmic horizons. This was true for our proposed hyperbolic ``black hole" geometry, and it is also true for solutions given by patches of the Reissner--Nordstr\"om black hole inside the inner horizon \cite{upcoming}. Negative $K$ also appears in calculations of conformal boundary conditions in de Sitter space \cite{Anninos:2024wpy}, which of course comes equipped with a cosmic horizon. This negativity of $K$ for cosmic horizons bears some similarity to the purported relation between AdS/CFT and dS/CFT as an analytic continuation in the number of degrees of freedom (with $N \rightarrow -N$ as long as $N \sim (\ell_{\text{AdS}}/\ell_{\text{Pl}})^{4n-2}$ for $n \in \mathbb{Z}^+$), made most precise in \cite{Anninos:2011ui}. 

\section*{Acknowledgments}
The authors would like to thank Dionysios Anninos and Eva Silverstein for fruitful conversations. The authors are supported in part by DOE grant DE-SC001010 and the Federico and Elvia Faggin Foundation.
BB performed this work in part during a visit to Aspen Center for Physics, which is supported by National Science Foundation grant PHY-2210452 and a grant from the Simons Foundation (1161654, Troyer).

\appendix
\section{Conformal Hamiltonian and energy}\label{appA}
The Hamiltonian in general relativity, like in any other theory, can be obtained from the Legendre transform of the Lagrangian. We begin with the Lorentzian action 
\be \label{I-alpha}
I = \frac{1}{16\pi G}\int_{\mathcal{M}} \sqrt{-g}\,R + \frac{\alpha}{8\pi G} \int_{\partial\mathcal{M}} \sqrt{-h}\,K,
\ee
where $\alpha=1$ and $\alpha = 1/d$ for Dirichlet and conformal boundary conditions, respectively. In the following we will decompose the action in canonical variables to implement the Legendre transform and track the boundary terms; for further details about canonical decomposition see e.g.~appendix E of \cite{Wald:1984rg} or \cite{Hawking:1995fd}.

We foliate the spacetime $\mc{M}$ by spatial slices $\Sigma$ with unit normal $u^\mu$ that are parameterized by the time coordinate $t$. The standard ADM decomposition of the line element is
\be
ds^2=-N^2 dt^2 + q_{ij}(N^i dt + dx^i)(N^j dt +dx^j) ,
\ee
where $N$ is the lapse function, $N^i$ is the shift vector and $q_{ij}$ is the spatial metric; space indices are denoted by Latin letters.
The canonical variables are the metric components $q_{ij}$ and their conjugate momenta $p^{ij}$.

The Ricci scalar in \eqref{I-alpha} can be written as 
\be
R=2 (G_{\mu \nu}-R_{\mu\nu})u^\mu u^\nu ,
\ee
where $G_{\mu\nu}$ is the Einstein tensor. The first term above
is a bulk integral that vanishes when the Hamiltonian constraint is satisfied,
while the second term can be written as
\be \label{Ruu}
R_{\mu\nu}u^\mu u^\nu = \mc{K}^2 -\mc{K}_{ij}\mc{K}^{ij} -\nabla_\nu (u^\nu \nabla_\mu u^\mu) +\nabla_\mu (u^\nu \nabla_\nu u^\mu).
\ee
$\mc{K}_{ij}$ is the extrinsic curvature of the constant-time slice and $\mc{K}$ is its trace. 
Using relations 
\be
\sqrt{-g}=N \sqrt{q} \quad , \quad 
2N\mc{K}_{ij}=\dot{q}_{ij}-2 D_{(i}N_{j)} 
 \quad , \quad 
16\pi G \, p^{ij}=\sqrt{q} (\mc{K}^{ij}- \mc{K}q^{ij})
\ee
where $D_i$ is the spatial covariant derivative compatible with $q_{ij}$, we have
\be \label{int-KK}
\int_{\mc{M}}\sqrt{-g} \, (\mc{K}_{ij}\mc{K}^{ij}- \mc{K}^2) = 8\pi G \int_{\mc{M}} \left ( p^{ij } \dot{q}_{ij} + 2N_{i}D_{j}p^{ij} \right ) -\int_{\partial \mc{M}} dS_i N_j (\mc{K}^{ij}- \mc{K}q^{ij}).
\ee
In the last term, $dS_i$ is the directed surface element of the codimension-2 intersection of the spatial slice with the timelike boundary $\partial \mc{M}$. The integral of the terms quadratic in extrinsic curvature in \eqref{Ruu} is thus a boundary integral involving the boundary value of the shift vector; it has no contribution in a gauge where the boundary shift is zero, which is what we pick now. Moreover, the $D_j p^{ij }$ term in \eqref{int-KK} vanishes when the momentum constraint is satisfied. 

The first total divergence in \eqref{Ruu} gives zero on the boundary $\partial \mc{M}$ with unit normal $n^\mu$, upon picking $u \cdot n =0 $ on the boundary. Using $u \cdot n =0$ lets us write the final term on the system boundary as\footnote{If there is a horizon, it can be shown that the same term on a smooth horizon produces the ``area'' term $-2\pi A_h^{(d-1)}$.} 
\be
\int_{\partial \mc{M}}\! \sqrt{-h} \, u^\mu u^\nu \nabla_\mu n_\nu = \int_{\partial \mc{M}}\! \sqrt{-h} \, \left [ (u^\nu u^\nu + g^{\mu\nu}) - g^{\mu\nu} \right ] \nabla_\mu n_\nu = \int dt \int d^{d-1}x \, N \sqrt{\sigma_{d-1}} \, (k-K). 
\ee
We used $(g^{\mu\nu} + u^\mu u^\nu) \nabla_\mu n_\nu = k $, where $k$ is the trace of extrinsic curvature of the $(d-1)$-dimensional boundary spatial slice as embedded in $\Sigma$. The metric determinant of this slice is denoted by $\sigma_{d-1}$.
Including the last term in \eqref{I-alpha} and writing the action as $\int dt (\int p^{ij}\dot{q}_{ij}-H)$, we may thus read off the Hamiltonian. It is a sum of bulk integrals which vanish when the constraints are satisfied, and a boundary term which we will call the energy $E$, since it equals $H$ in the constrained phase space:
\be\label{Enformal}
E= -\frac{1}{8\pi G}\int d^{d-1}x \, N \sqrt{\sigma_{d-1}} \, [k+(\alpha -1)K].
\ee
Note that for the Dirichlet case ($\a = 1$) we get the familiar ADM expression (or the Brown--York energy, when the boundary lapse is 1), modulo any counterterms. This Hamiltonian was previously obtained in \cite{Odak:2021axr}. 

For a spacetime with the metric,
\be \label{eq:spherical-metric}
ds^2=-f(r)dt^2+f^{-1}(r)dr^2+ r^2 d\Omega_{d-1}^2,
\ee
that is bounded by a boundary at $r=r_c$, we have $k=(d-1)\sqrt{f(r_c)}/r_c$, and $K$ is given in \eqref{indmetk}. Thus the energy with conformal boundary conditions becomes
\be \label{H-CBC}
E^{\rm{CBC}}=- N_\partial \frac{\text{Vol}[S^{d-1}]}{8\pi G} \frac{d-1}{d}r_c^{d-2}\sqrt{f(r_c)}\left (1-\frac{r_c}{2}\frac{f'(r_c)}{f(r_c)} \right),
\ee
where $N|_{\partial \mathcal{M}} = N_{\partial}$ and we use CBC as a shorthand to highlight the choice of conformal boundary conditions. The outward normal vector to the system boundary points toward larger spheres. In order to get the value of the Hamiltonian generating the boundary {\it proper} time, we drop the boundary lapse factor.

It is also instructive to calculate something analogous to the Brown--York stress tensor in this context. Varying the action \eqref{I-alpha} gives
\be\label{Ivary}
(8\pi G)\delta I = \frac{1}{2}\int_{\mc{M}}\sqrt{-g}\,G_{\mu\nu}\delta g^{\mu\nu}+\frac{1}{2}\int_{\partial \mc{M}}\sqrt{-h} \left ( K_{\mu\nu}-\alpha K h_{\mu\nu} \right )\delta h^{\mu\nu} + \int_{\partial \mc{M}}\sqrt{-h} (\alpha -1) \delta K.
\ee
Defining the stress tensor with conformal boundary conditions by variation of the on-shell action with respect to the boundary metric (with $K$ held fixed) gives
\be\label{T-CBC}
T_{\mu\nu}^{\rm CBC}=-\frac{2}{\sqrt{-h}}\frac{\delta I}{\delta h^{\mu\nu}}= - \frac{1}{8\pi G}\left(K_{\mu\nu}-\frac{1}{d}Kh_{\mu\nu}\right),
\ee
and the energy associated to the configuration \eqref{eq:spherical-metric} is 
\be\label{E-CBCApp}
E^{\rm CBC}= \int \! d^{d-1}x \, \sqrt{\sigma_{d-1}} \, u^\mu u^\nu T_{\mu\nu}^{\rm CBC}= - \frac{\text{Vol}[S^{d-1}]}{8\pi G} \frac{d-1}{d}r_c^{d-2}\sqrt{f(r_c)}\left (1-\frac{r_c}{2}\frac{f'(r_c)}{f(r_c)} \right).
\ee
This matches the above result \eqref{H-CBC} from the Hamiltonian computation, up to the boundary lapse (which equals $1$ when $t$ becomes the proper time on the boundary).

It is natural to question the use of the Brown--York stress tensor in a situation where we are fixing the conformal class of metrics instead of fixing a precise metric. Another option is to define a modified Brown--York stress ``tensor" where we vary with respect to the conformal metric $\overline{h}_{\m\n} = (-h)^{-1/d}h_{\m\n}$. This gives $\d h^{\m\n} = (-h)^{-1/d} \d \overline{h}^{\m\n} + \f 1 d \overline{h}^{\m\n} (-h)^{-1/d-1} \d h$, so we have 
\be
\frac{1}{2}\int_{\partial \mc{M}}\sqrt{-h} \left ( K_{\mu\nu}-\frac{1}{d} K h_{\mu\nu} \right )\delta h^{\mu\nu} = \frac{1}{2}\int_{\partial \mc{M}}\sqrt{-h}\,(-h)^{-1/d} \left ( K_{\mu\nu}-\frac{1}{d} K h_{\mu\nu} \right )\delta \overline{h}^{\mu\nu} .
\ee
Therefore, defining the stress tensor by varying with respect to the conformal metric gives (recall $\overline{h}=-1$)
\be
\overline{T}^{\rm CBC}_{\mu\nu }= -\frac{2}{\sqrt{-\overline{h}}}\frac{\delta I}{\delta \overline{h}^{\mu\nu}} = - \frac{1}{8\pi G}\sqrt{-h}\,(-h)^{-1/d} \left ( K_{\mu\nu}-\frac{1}{d} K h_{\mu\nu} \right ),
\ee
which is different from the above definition by factors of $h$. Interestingly, this object remains traceless. However, since the conformal metric $\overline{h}_{\m\n}$ is not a proper tensor (but rather a tensor density), the object above is also a stress tensor density instead of an ordinary stress tensor.

It is important to note when comparing to the main text that the thermal energy $\tilde{E}$ calculated from the partition function by $\tilde{E}=\partial_{\tilde{\b}}I$ (as in \cite{Anninos:2023epi}) differs by a factor of $r_c$; namely it satisfies $\tilde{E} = r_c E^{\rm CBC} $. Thus $\tilde{\beta} \tilde{E}=\beta E^{\rm CBC}$. This is expected since in the definition of the partition function \eqref{Z-conformal} we have $\tilde{\beta}\tilde{E}$ in the exponent.

\section{Exact solutions in 5d bulk}\label{app:exact5d}
Exact solutions are possible in both $d=3$ and $d = 4$. The former was studied by \cite{Anninos:2023epi}, so let us focus on the latter. For $d = 4$ on a spatial sphere we have 
\be\label{sphericalexact}
r_h = \frac{\pi\Big(\pi \pm \sqrt{\pi ^2 - \tilde{\beta}^2}\Big) + \tilde{\beta}^2}{\pi \tilde{\beta} K},\qquad r_c = \sqrt{\frac{2}{\pi}} \frac{\Big(2\pi \mp \sqrt{\pi^2 - \tilde{\beta}^2}\Big)\sqrt{\pi \pm \sqrt{\pi^2 - \tilde{\beta}^2}}}{\tilde{\beta}K}.
\ee
The upper signs correspond to the large black hole while the lower signs correspond to the small black hole. We only have black holes for $\tilde{\b} \leq \pi$, and we always have $r_c \geq r_h$.

\begin{figure}
\centering
\includegraphics[scale=0.7]{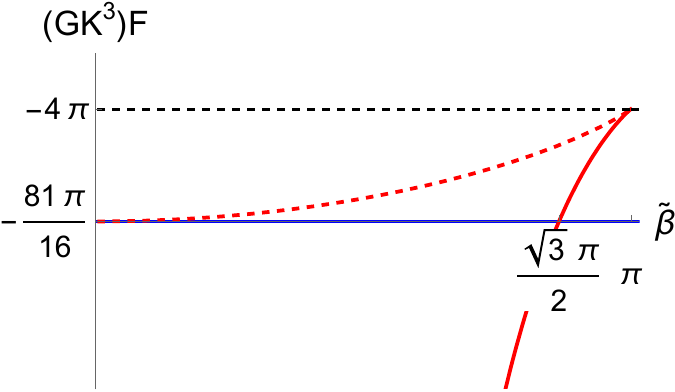}
\caption{The free energy diagram showing $GK^3F$ as a function of inverse conformal temperature $\tilde{\beta}$, as written in \eqref{freeenergy}. The blue line is the Casimir energy, i.e.~the free energy of the thermal gas. The red dashed line represents the small black hole, and the red solid line represents the large black hole. Both branches meet at the critical point $\tilde{\beta} = \pi$, but for $\tilde{\beta} < \pi$ the small black hole has the highest free energy and thus is subdominant in the ensemble. By contrast, the large black hole is dominant for temperatures above the Hawking--Page transition at $\tilde{\beta} = \sqrt{3}\pi/2$.}
\label{figs:freeEnergyDiagram}
\end{figure}

As in $d = 3$, we have a Hawking--Page transition in the free energy $F \equiv -\tilde{\beta}^{-1}\log Z$. Applying a saddlepoint approximation to the gravitational path integral and recalling that the volume of a unit $S^3$ is $2\pi^2$, we have that $F = -\pi K r_c^4/(16 G)$ and thus 
\begin{equation}
F_{\text{black hole}} = -\f{\left(\pi^3 + 3\pi \tilde{\b}^2 \pm (\pi^2 - \tilde{\b}^2)^{3/2}\right)^2}{4 \pi G K^3 \tilde{\b}^4},\qquad F_{\text{thermal gas}} = -\dfrac{81\pi}{16 GK^3}.\label{freeenergy}
\end{equation}
We have plotted the free-energy phase diagram in Figure \ref{figs:freeEnergyDiagram}. It is qualitatively the same as both the $d = 3$ case \cite{Anninos:2023epi} and AdS space on $S^1 \times S^{d-1}$ with an asymptotic Dirichlet boundary \cite{Hawking:1982dh,Witten:1998zw}. Here, the critical temperature below which there are no black hole solutions is given by $\tilde{\beta}_{\text{max}} = \pi$, obtained by plugging $d=4$ into \eqref{betamax}. The Hawking--Page temperature is
\begin{equation}
\tilde{\beta}_{\text{HP}} = \frac{\sqrt{3}\pi}{2}.
\end{equation}
We can also write an exact solution for the boundary theory on hyperbolic space $\mathbb{H}^3$: 
\be
r_h = \frac{\pi\Big(\pi + \sqrt{\pi ^2 + \tilde{\beta}^2}\Big) -\tilde{\beta}^2}{\pi \tilde{\beta} K}, \qquad r_c = \sqrt{\f{2}{\pi}}\f{\Big(2\pi - \sqrt{\pi^2 + \tilde{\beta}^2}\Big)\sqrt{\pi + \sqrt{\pi^2 + \tilde{\beta}^2}}}{\tilde{\b}K}.
\ee
As discussed in the main text, there is only one solution for a given range of parameters. Notice that $r_c$ and $r_h$ go to zero as $\tilde{\b} \rightarrow \sqrt{3}\pi$, which is an indication that the geometric description is breaking down. The free energy is given by 
\be
F = -\f{\text{Vol}(\mathbb{H}^3)\left(\pi^3 - 3\pi \tilde{\b}^2 + (\pi^2 + \tilde{\b}^2)^{3/2}\right)^2}{8 \pi^3 G K^3 \tilde{\b}^4}.
\ee
As argued around \eqref{bdrydatarelation}, the hyperbolic formulas for $r_h$, $r_c$, and $-\log Z = \tilde{\beta}F$ can be obtained from those of the large spherical black hole by taking $K \rightarrow iK$, $\tilde{\b} \rightarrow - i \tilde{\b}$. Expanding $F$ of the spherical and hyperbolic black holes at high temperature, where they dominate their respective ensembles, gives the requisite sign flips in every other term in the thermal effective action:
\be\label{tefthyperbolic}
K>0:\quad -\log Z = -\f{1}{G K^3} \left(\f{\pi^5}{\tilde{\b}^3} \pm \f{3\pi^3}{2\tilde{\b}} + \f{15 \pi \tilde{\b}}{16} \pm \f{11 \tilde{\b}^3}{32\pi} + \cdots \right).
\ee
The upper signs correspond to the action on $S^3$ whereas the lower signs correspond to the action on $\mathbb{H}^3$. We have set Vol$(\mathbb{H}^3) = 2\pi^2$ for an accurate comparison to the spherical case.

\small
\bibliographystyle{jhep}
\bibliography{references.bib}

\providecommand{\href}[2]{#2}\begingroup\raggedright\begin{thebibliography}{10}

\bibitem{tHooft:1993dmi}
G.~'t~Hooft, \emph{{Dimensional reduction in quantum gravity}}, {\emph{Conf.
  Proc. C} {\bfseries 930308} (1993) 284}
  [\href{https://arxiv.org/abs/gr-qc/9310026}{{\ttfamily gr-qc/9310026}}].

\bibitem{Susskind:1994vu}
L.~Susskind, \emph{{The World as a hologram}},
  \href{https://doi.org/10.1063/1.531249}{\emph{J. Math. Phys.} {\bfseries 36}
  (1995) 6377} [\href{https://arxiv.org/abs/hep-th/9409089}{{\ttfamily
  hep-th/9409089}}].

\bibitem{McGough:2016lol}
L.~McGough, M.~Mezei and H.~Verlinde, \emph{{Moving the CFT into the bulk with
  $ T\overline{T} $}},
  \href{https://doi.org/10.1007/JHEP04(2018)010}{\emph{JHEP} {\bfseries 04}
  (2018) 010} [\href{https://arxiv.org/abs/1611.03470}{{\ttfamily
  1611.03470}}].

\bibitem{Taylor:2018xcy}
M.~Taylor, \emph{{$T \bar{T}$ deformations in general dimensions}},
  \href{https://doi.org/10.4310/ATMP.2023.v27.n1.a2}{\emph{Adv. Theor. Math.
  Phys.} {\bfseries 27} (2023) 37}
  [\href{https://arxiv.org/abs/1805.10287}{{\ttfamily 1805.10287}}].

\bibitem{Hartman:2018tkw}
T.~Hartman, J.~Kruthoff, E.~Shaghoulian and A.~Tajdini, \emph{{Holography at
  finite cutoff with a $T^2$ deformation}},
  \href{https://doi.org/10.1007/JHEP03(2019)004}{\emph{JHEP} {\bfseries 03}
  (2019) 004} [\href{https://arxiv.org/abs/1807.11401}{{\ttfamily
  1807.11401}}].

\bibitem{Gorbenko:2018oov}
V.~Gorbenko, E.~Silverstein and G.~Torroba, \emph{{dS/dS and $ T\overline{T}
  $}}, \href{https://doi.org/10.1007/JHEP03(2019)085}{\emph{JHEP} {\bfseries
  03} (2019) 085} [\href{https://arxiv.org/abs/1811.07965}{{\ttfamily
  1811.07965}}].

\bibitem{Gross:2019ach}
D.~J. Gross, J.~Kruthoff, A.~Rolph and E.~Shaghoulian, \emph{{$T\overline{T}$
  in AdS$_2$ and Quantum Mechanics}},
  \href{https://doi.org/10.1103/PhysRevD.101.026011}{\emph{Phys. Rev. D}
  {\bfseries 101} (2020) 026011}
  [\href{https://arxiv.org/abs/1907.04873}{{\ttfamily 1907.04873}}].

\bibitem{Lewkowycz:2019xse}
A.~Lewkowycz, J.~Liu, E.~Silverstein and G.~Torroba, \emph{{$ T\overline{T} $
  and EE, with implications for (A)dS subregion encodings}},
  \href{https://doi.org/10.1007/JHEP04(2020)152}{\emph{JHEP} {\bfseries 04}
  (2020) 152} [\href{https://arxiv.org/abs/1909.13808}{{\ttfamily
  1909.13808}}].

\bibitem{Gross:2019uxi}
D.~J. Gross, J.~Kruthoff, A.~Rolph and E.~Shaghoulian, \emph{{Hamiltonian
  deformations in quantum mechanics, $T\bar T$, and the SYK model}},
  \href{https://doi.org/10.1103/PhysRevD.102.046019}{\emph{Phys. Rev. D}
  {\bfseries 102} (2020) 046019}
  [\href{https://arxiv.org/abs/1912.06132}{{\ttfamily 1912.06132}}].

\bibitem{Coleman:2020jte}
E.~Coleman and V.~Shyam, \emph{{Conformal boundary conditions from cutoff
  AdS$_{3}$}}, \href{https://doi.org/10.1007/JHEP09(2021)079}{\emph{JHEP}
  {\bfseries 09} (2021) 079}
  [\href{https://arxiv.org/abs/2010.08504}{{\ttfamily 2010.08504}}].

\bibitem{Coleman:2021nor}
E.~Coleman, E.~A. Mazenc, V.~Shyam, E.~Silverstein, R.~M. Soni, G.~Torroba and
  S.~Yang, \emph{{De Sitter microstates from T$ \overline{T} $ +
  \ensuremath{\Lambda}$_{2}$ and the Hawking-Page transition}},
  \href{https://doi.org/10.1007/JHEP07(2022)140}{\emph{JHEP} {\bfseries 07}
  (2022) 140} [\href{https://arxiv.org/abs/2110.14670}{{\ttfamily
  2110.14670}}].

\bibitem{Svesko:2022txo}
A.~Svesko, E.~Verheijden, E.~P. Verlinde and M.~R. Visser, \emph{{Quasi-local
  energy and microcanonical entropy in two-dimensional nearly de Sitter
  gravity}}, \href{https://doi.org/10.1007/JHEP08(2022)075}{\emph{JHEP}
  {\bfseries 08} (2022) 075}
  [\href{https://arxiv.org/abs/2203.00700}{{\ttfamily 2203.00700}}].

\bibitem{Banihashemi:2022jys}
B.~Banihashemi and T.~Jacobson, \emph{{Thermodynamic ensembles with
  cosmological horizons}},
  \href{https://doi.org/10.1007/JHEP07(2022)042}{\emph{JHEP} {\bfseries 07}
  (2022) 042} [\href{https://arxiv.org/abs/2204.05324}{{\ttfamily
  2204.05324}}].

\bibitem{Anninos:2023epi}
D.~Anninos, D.~A. Galante and C.~Maneerat, \emph{{Gravitational
  observatories}}, \href{https://doi.org/10.1007/JHEP12(2023)024}{\emph{JHEP}
  {\bfseries 12} (2023) 024}
  [\href{https://arxiv.org/abs/2310.08648}{{\ttfamily 2310.08648}}].

\bibitem{Anninos:2024wpy}
D.~Anninos, D.~A. Galante and C.~Maneerat, \emph{{Cosmological observatories}},
  \href{https://doi.org/10.1088/1361-6382/ad5824}{\emph{Class. Quant. Grav.}
  {\bfseries 41} (2024) 165009}
  [\href{https://arxiv.org/abs/2402.04305}{{\ttfamily 2402.04305}}].

\bibitem{Batra:2024kjl}
G.~Batra, G.~B. De~Luca, E.~Silverstein, G.~Torroba and S.~Yang,
  \emph{{Bulk-local dS$_3$ holography: the Matter with $T\bar T+\Lambda_2$}},
  \href{https://arxiv.org/abs/2403.01040}{{\ttfamily 2403.01040}}.

\bibitem{Batra:2024qju}
G.~Batra, \emph{{Timelike boundaries in de Sitter JT gravity and the Gao-Wald
  theorem}},  \href{https://arxiv.org/abs/2407.08913}{{\ttfamily 2407.08913}}.

\bibitem{Anderson:2006lqb}
M.~T. Anderson, \emph{{On boundary value problems for Einstein metrics}},
  \href{https://doi.org/10.2140/gt.2008.12.2009}{\emph{Geom. Topol.} {\bfseries
  12} (2008) 2009} [\href{https://arxiv.org/abs/math/0612647}{{\ttfamily
  math/0612647}}].

\bibitem{An:2021fcq}
Z.~An and M.~T. Anderson, \emph{{The initial boundary value problem and
  quasi-local Hamiltonians in General Relativity}},
  \href{https://doi.org/10.1088/1361-6382/ac0a86}{\emph{Classical and Quantum
  Gravity} {\bfseries 38} (2021) 154001}
  [\href{https://arxiv.org/abs/2103.15673}{{\ttfamily 2103.15673}}].

\bibitem{allameh}
K.~Allameh and E.~Shaghoulian, \emph{{Modular invariance and thermal effective
  field theory in CFT}},  \href{https://arxiv.org/abs/2402.13337}{{\ttfamily
  2402.13337}}.

\bibitem{Witten:2018lgb}
E.~Witten, \emph{{A note on boundary conditions in Euclidean gravity}},
  \href{https://doi.org/10.1142/S0129055X21400043}{\emph{Rev. Math. Phys.}
  {\bfseries 33} (2021) 2140004}
  [\href{https://arxiv.org/abs/1805.11559}{{\ttfamily 1805.11559}}].

\bibitem{Anninos:2022ujl}
D.~Anninos, D.~A. Galante and B.~M\"uhlmann, \emph{{Finite features of quantum
  de Sitter space}},
  \href{https://doi.org/10.1088/1361-6382/acaba5}{\emph{Class. Quant. Grav.}
  {\bfseries 40} (2023) 025009}
  [\href{https://arxiv.org/abs/2206.14146}{{\ttfamily 2206.14146}}].

\bibitem{Bredberg:2011xw}
I.~Bredberg and A.~Strominger, \emph{{Black Holes as Incompressible Fluids on
  the Sphere}}, \href{https://doi.org/10.1007/JHEP05(2012)043}{\emph{JHEP}
  {\bfseries 05} (2012) 043} [\href{https://arxiv.org/abs/1106.3084}{{\ttfamily
  1106.3084}}].

\bibitem{Anninos:2011zn}
D.~Anninos, T.~Anous, I.~Bredberg and G.~S. Ng, \emph{{Incompressible Fluids of
  the de Sitter Horizon and Beyond}},
  \href{https://doi.org/10.1007/JHEP05(2012)107}{\emph{JHEP} {\bfseries 05}
  (2012) 107} [\href{https://arxiv.org/abs/1110.3792}{{\ttfamily 1110.3792}}].

\bibitem{York:1972sj}
J.~W. York, Jr., \emph{{Role of conformal three geometry in the dynamics of
  gravitation}}, \href{https://doi.org/10.1103/PhysRevLett.28.1082}{\emph{Phys.
  Rev. Lett.} {\bfseries 28} (1972) 1082}.

\bibitem{York:1986lje}
J.~York, \emph{{Boundary terms in the action principles of general
  relativity}}, \href{https://doi.org/10.1007/BF01889475}{\emph{Found. Phys.}
  {\bfseries 16} (1986) 249}.

\bibitem{Odak:2021axr}
G.~Odak and S.~Speziale, \emph{{Brown-York charges with mixed boundary
  conditions}}, \href{https://doi.org/10.1007/JHEP11(2021)224}{\emph{JHEP}
  {\bfseries 11} (2021) 224}
  [\href{https://arxiv.org/abs/2109.02883}{{\ttfamily 2109.02883}}].

\bibitem{Jensen:2012jh}
K.~Jensen, M.~Kaminski, P.~Kovtun, R.~Meyer, A.~Ritz and A.~Yarom,
  \emph{{Towards hydrodynamics without an entropy current}},
  \href{https://doi.org/10.1103/PhysRevLett.109.101601}{\emph{Phys. Rev. Lett.}
  {\bfseries 109} (2012) 101601}
  [\href{https://arxiv.org/abs/1203.3556}{{\ttfamily 1203.3556}}].

\bibitem{Banerjee:2012iz}
N.~Banerjee, J.~Bhattacharya, S.~Bhattacharyya, S.~Jain, S.~Minwalla and
  T.~Sharma, \emph{{Constraints on Fluid Dynamics from Equilibrium Partition
  Functions}}, \href{https://doi.org/10.1007/JHEP09(2012)046}{\emph{JHEP}
  {\bfseries 09} (2012) 046} [\href{https://arxiv.org/abs/1203.3544}{{\ttfamily
  1203.3544}}].

\bibitem{horowitz}
G.~T. Horowitz and E.~Shaghoulian, \emph{{Detachable circles and
  temperature-inversion dualities for CFT$_{d}$}},
  \href{https://doi.org/10.1007/JHEP01(2018)135}{\emph{JHEP} {\bfseries 01}
  (2018) 135} [\href{https://arxiv.org/abs/1709.06084}{{\ttfamily
  1709.06084}}].

\bibitem{Benjamin:2023qsc}
N.~Benjamin, J.~Lee, H.~Ooguri and D.~Simmons-Duffin, \emph{{Universal
  asymptotics for high energy CFT data}},
  \href{https://doi.org/10.1007/JHEP03(2024)115}{\emph{JHEP} {\bfseries 03}
  (2024) 115} [\href{https://arxiv.org/abs/2306.08031}{{\ttfamily
  2306.08031}}].

\bibitem{Kutasov:2000td}
D.~Kutasov and F.~Larsen, \emph{{Partition sums and entropy bounds in weakly
  coupled CFT}},
  \href{https://doi.org/10.1088/1126-6708/2001/01/001}{\emph{JHEP} {\bfseries
  01} (2001) 001} [\href{https://arxiv.org/abs/hep-th/0009244}{{\ttfamily
  hep-th/0009244}}].

\bibitem{Melia:2020pzd}
T.~Melia and S.~Pal, \emph{{EFT Asymptotics: the Growth of Operator
  Degeneracy}},
  \href{https://doi.org/10.21468/SciPostPhys.10.5.104}{\emph{SciPost Phys.}
  {\bfseries 10} (2021) 104}
  [\href{https://arxiv.org/abs/2010.08560}{{\ttfamily 2010.08560}}].

\bibitem{Liu:2024ymn}
X.~Liu, J.~E. Santos and T.~Wiseman, \emph{{New Well-Posed Boundary Conditions
  for Semi-Classical Euclidean Gravity}},
  \href{https://arxiv.org/abs/2402.04308}{{\ttfamily 2402.04308}}.

\bibitem{Geroch:1982bv}
R.~P. Geroch and J.~B. Hartle, \emph{{Distorted black holes}},
  \href{https://doi.org/10.1063/1.525384}{\emph{J. Math. Phys.} {\bfseries 23}
  (1982) 680}.

\bibitem{Hawking:1982dh}
S.~W. Hawking and D.~N. Page, \emph{{Thermodynamics of Black Holes in anti-De
  Sitter Space}}, \href{https://doi.org/10.1007/BF01208266}{\emph{Commun. Math.
  Phys.} {\bfseries 87} (1983) 577}.

\bibitem{Witten:1998zw}
E.~Witten, \emph{{Anti-de Sitter space, thermal phase transition, and
  confinement in gauge theories}},
  \href{https://doi.org/10.4310/ATMP.1998.v2.n3.a3}{\emph{Adv. Theor. Math.
  Phys.} {\bfseries 2} (1998) 505}
  [\href{https://arxiv.org/abs/hep-th/9803131}{{\ttfamily hep-th/9803131}}].

\bibitem{Horowitz:1998ha}
G.~T. Horowitz and R.~C. Myers, \emph{{The AdS / CFT correspondence and a new
  positive energy conjecture for general relativity}},
  \href{https://doi.org/10.1103/PhysRevD.59.026005}{\emph{Phys. Rev. D}
  {\bfseries 59} (1998) 026005}
  [\href{https://arxiv.org/abs/hep-th/9808079}{{\ttfamily hep-th/9808079}}].

\bibitem{Belin:2016yll}
A.~Belin, J.~de~Boer, J.~Kruthoff, B.~Michel, E.~Shaghoulian and M.~Shyani,
  \emph{{Universality of sparse $d > 2$ conformal field theory at large $N$}},
  \href{https://doi.org/10.1007/JHEP03(2017)067}{\emph{JHEP} {\bfseries 03}
  (2017) 067} [\href{https://arxiv.org/abs/1610.06186}{{\ttfamily
  1610.06186}}].

\bibitem{Hawking:1971vc}
S.~W. Hawking, \emph{{Black holes in general relativity}},
  \href{https://doi.org/10.1007/BF01877517}{\emph{Commun. Math. Phys.}
  {\bfseries 25} (1972) 152}.

\bibitem{Kubiznak:2016qmn}
D.~Kubiznak, R.~B. Mann and M.~Teo, \emph{{Black hole chemistry: thermodynamics
  with Lambda}}, \href{https://doi.org/10.1088/1361-6382/aa5c69}{\emph{Class.
  Quant. Grav.} {\bfseries 34} (2017) 063001}
  [\href{https://arxiv.org/abs/1608.06147}{{\ttfamily 1608.06147}}].

\bibitem{York:1986it}
J.~W. York, Jr., \emph{{Black hole thermodynamics and the Euclidean Einstein
  action}}, \href{https://doi.org/10.1103/PhysRevD.33.2092}{\emph{Phys. Rev. D}
  {\bfseries 33} (1986) 2092}.

\bibitem{Casini:2011kv}
H.~Casini, M.~Huerta and R.~C. Myers, \emph{{Towards a derivation of
  holographic entanglement entropy}},
  \href{https://doi.org/10.1007/JHEP05(2011)036}{\emph{JHEP} {\bfseries 05}
  (2011) 036} [\href{https://arxiv.org/abs/1102.0440}{{\ttfamily 1102.0440}}].

\bibitem{Zamolodchikov:1986gt}
A.~B. Zamolodchikov, \emph{{Irreversibility of the Flux of the Renormalization
  Group in a 2D Field Theory}}, {\emph{JETP Lett.} {\bfseries 43} (1986) 730}.

\bibitem{Casini:2004bw}
H.~Casini and M.~Huerta, \emph{{A Finite entanglement entropy and the
  c-theorem}},
  \href{https://doi.org/10.1016/j.physletb.2004.08.072}{\emph{Phys. Lett. B}
  {\bfseries 600} (2004) 142}
  [\href{https://arxiv.org/abs/hep-th/0405111}{{\ttfamily hep-th/0405111}}].

\bibitem{Komargodski:2011vj}
Z.~Komargodski and A.~Schwimmer, \emph{{On Renormalization Group Flows in Four
  Dimensions}}, \href{https://doi.org/10.1007/JHEP12(2011)099}{\emph{JHEP}
  {\bfseries 12} (2011) 099} [\href{https://arxiv.org/abs/1107.3987}{{\ttfamily
  1107.3987}}].

\bibitem{Casini:2012ei}
H.~Casini and M.~Huerta, \emph{{On the RG running of the entanglement entropy
  of a circle}}, \href{https://doi.org/10.1103/PhysRevD.85.125016}{\emph{Phys.
  Rev. D} {\bfseries 85} (2012) 125016}
  [\href{https://arxiv.org/abs/1202.5650}{{\ttfamily 1202.5650}}].

\bibitem{Casini:2017vbe}
H.~Casini, E.~Test\'e and G.~Torroba, \emph{{Markov Property of the Conformal
  Field Theory Vacuum and the a Theorem}},
  \href{https://doi.org/10.1103/PhysRevLett.118.261602}{\emph{Phys. Rev. Lett.}
  {\bfseries 118} (2017) 261602}
  [\href{https://arxiv.org/abs/1704.01870}{{\ttfamily 1704.01870}}].

\bibitem{Fisher:1986zz}
D.~S. Fisher, \emph{{Scaling and critical slowing down in random-field Ising
  systems}}, \href{https://doi.org/10.1103/PhysRevLett.56.416}{\emph{Phys. Rev.
  Lett.} {\bfseries 56} (1986) 416}.

\bibitem{Ogawa:2011bz}
N.~Ogawa, T.~Takayanagi and T.~Ugajin, \emph{{Holographic Fermi Surfaces and
  Entanglement Entropy}},
  \href{https://doi.org/10.1007/JHEP01(2012)125}{\emph{JHEP} {\bfseries 01}
  (2012) 125} [\href{https://arxiv.org/abs/1111.1023}{{\ttfamily 1111.1023}}].

\bibitem{Huijse:2011ef}
L.~Huijse, S.~Sachdev and B.~Swingle, \emph{{Hidden Fermi surfaces in
  compressible states of gauge-gravity duality}},
  \href{https://doi.org/10.1103/PhysRevB.85.035121}{\emph{Phys. Rev. B}
  {\bfseries 85} (2012) 035121}
  [\href{https://arxiv.org/abs/1112.0573}{{\ttfamily 1112.0573}}].

\bibitem{Emparan:1999pm}
R.~Emparan, C.~V. Johnson and R.~C. Myers, \emph{{Surface terms as counterterms
  in the AdS / CFT correspondence}},
  \href{https://doi.org/10.1103/PhysRevD.60.104001}{\emph{Phys. Rev. D}
  {\bfseries 60} (1999) 104001}
  [\href{https://arxiv.org/abs/hep-th/9903238}{{\ttfamily hep-th/9903238}}].

\bibitem{upcoming}
B.~Banihashemi, E.~Shaghoulian and S.~Shashi, \emph{{work in progress}}, .

\bibitem{Adam:2011dn}
A.~Adam, S.~Kitchen and T.~Wiseman, \emph{{A numerical approach to finding
  general stationary vacuum black holes}},
  \href{https://doi.org/10.1088/0264-9381/29/16/165002}{\emph{Class. Quant.
  Grav.} {\bfseries 29} (2012) 165002}
  [\href{https://arxiv.org/abs/1105.6347}{{\ttfamily 1105.6347}}].

\bibitem{Sahakian:1999bd}
V.~Sahakian, \emph{{Holography, a covariant c function, and the geometry of the
  renormalization group}},
  \href{https://doi.org/10.1103/PhysRevD.62.126011}{\emph{Phys. Rev. D}
  {\bfseries 62} (2000) 126011}
  [\href{https://arxiv.org/abs/hep-th/9910099}{{\ttfamily hep-th/9910099}}].

\bibitem{Girardello:1998pd}
L.~Girardello, M.~Petrini, M.~Porrati and A.~Zaffaroni, \emph{{Novel local CFT
  and exact results on perturbations of N=4 superYang Mills from AdS
  dynamics}}, \href{https://doi.org/10.1088/1126-6708/1998/12/022}{\emph{JHEP}
  {\bfseries 12} (1998) 022}
  [\href{https://arxiv.org/abs/hep-th/9810126}{{\ttfamily hep-th/9810126}}].

\bibitem{Freedman:1999gp}
D.~Z. Freedman, S.~S. Gubser, K.~Pilch and N.~P. Warner, \emph{{Renormalization
  group flows from holography supersymmetry and a c theorem}},
  \href{https://doi.org/10.4310/ATMP.1999.v3.n2.a7}{\emph{Adv. Theor. Math.
  Phys.} {\bfseries 3} (1999) 363}
  [\href{https://arxiv.org/abs/hep-th/9904017}{{\ttfamily hep-th/9904017}}].

\bibitem{Anninos:2011ui}
D.~Anninos, T.~Hartman and A.~Strominger, \emph{{Higher Spin Realization of the
  dS/CFT Correspondence}},
  \href{https://doi.org/10.1088/1361-6382/34/1/015009}{\emph{Class. Quant.
  Grav.} {\bfseries 34} (2017) 015009}
  [\href{https://arxiv.org/abs/1108.5735}{{\ttfamily 1108.5735}}].

\bibitem{Wald:1984rg}
R.~M. Wald, \emph{{General Relativity}}. Chicago Univ. Pr., Chicago, USA, 1984,
  \href{https://doi.org/10.7208/chicago/9780226870373.001.0001}{10.7208/chicago/9780226870373.001.0001}.

\bibitem{Hawking:1995fd}
S.~W. Hawking and G.~T. Horowitz, \emph{{The Gravitational Hamiltonian, action,
  entropy and surface terms}},
  \href{https://doi.org/10.1088/0264-9381/13/6/017}{\emph{Class. Quant. Grav.}
  {\bfseries 13} (1996) 1487}
  [\href{https://arxiv.org/abs/gr-qc/9501014}{{\ttfamily gr-qc/9501014}}].

\end{thebibliography}\endgroup

\end{document}